\documentstyle[12pt, aaspp4, tighten, flushrt, psfig]{article}
\begin{document}

\title{Emission-Line  Properties of the Optical Filaments of NGC~1275}
\author{Bassem M. Sabra \& Joseph C. Shields}
\affil{\it Department of Physics \& Astronomy, Ohio University, 
Athens, OH 45701}
\and
\author{Alexei V. Filippenko}
\affil{\it Department of Astronomy, University of California, 
Berkeley, CA 94720-3411}

\begin{abstract}
Extended nebular filaments are seen at optical wavelengths in 
NGC~1275, the central galaxy in the Perseus cluster.  The agents
responsible for the excitation of these filaments remain poorly
understood.  In this paper we investigate possible mechanisms for
powering the filaments, using measurements from an extensive
spectroscopic data set acquired at the Lick Observatory 3-m Shane
telescope.  The results show that the filaments are in an extremely
low ionization and excitation state. The high signal-to-noise ratio
of the spectra allows us to measure or place sensitive upper limits on
weak but important diagnostic lines. We compare the observed line
intensity ratios to the predictions of various ionization models, including
photoionization by an active galactic nucleus, shock heating, stellar 
photoionization, and
photoionization by the intracluster medium.  We also investigate 
possible roles for cluster extreme-ultraviolet emission, and filtering
of cluster soft X-ray emission by an ionized screen, in the energetics
of the filaments.  None of these mechanisms provides an entirely
satisfactory explanation for the physical state of the nebulae.
Heating and ionization by reconnection of the intracluster magnetic
field remains a potentially viable alternative, which merits further
investigation through Faraday rotation studies.
\end{abstract}


\section{\sc Introduction}
Extended nebular filaments are commonly seen in association with the
central galaxies of galaxy clusters.  Clusters that feature a massive
central galaxy are also often identified as having a cooling flow,
based on cooling-timescale arguments or other evidence (Fabian, Nulsen, 
\& Canizares 1984).
This association has prompted speculation that the optical nebulosity 
is linked causally or energetically to the intracluster medium (ICM), and 
hence may be an important diagnostic of the cooling-flow phenomenon. 
As a result, several previous studies have investigated optical 
``cooling-flow filaments'' in attempts to understand their physical 
properties.  These efforts have helped delineate the observed 
characteristics of the nebulosity, and have prompted a variety of 
theoretical explanations for what is seen. 

Optical spectroscopy reveals that the filaments emit strongly in 
low-ionization forbidden lines, thus resembling Low Ionization Nuclear 
Emission-Line Regions (LINERs; Heckman 1980).  Heckman et al. (1989) also 
reported evidence that cooling-flow filaments can be separated into 
two characteristic groups based on their line-intensity ratios.  Typical
ratios for Type I systems are 
[\ion{N}{2}]$\lambda6583$/H$\alpha \approx 2$,
[\ion{S}{2}]$\lambda6717$/H$\alpha \approx 0.7$, and
[\ion{O}{1}]$\lambda6300$/H$\alpha \approx 0.2$;
these objects tend to have  
low X-ray and H$\alpha$ luminosities and are associated with low cooling 
rates and small optical filaments.  In contrast, typical ratios for
Type II objects are
[\ion{N}{2}]$\lambda6583$/H$\alpha \approx 1$,
[\ion{S}{2}]$\lambda6717$/H$\alpha \approx 0.4$, and
[\ion{O}{1}]$\lambda6300$/H$\alpha \approx 0.3$;
these sources have high X-ray and 
H$\alpha$ luminosities and are associated with high inferred cooling 
rates and extended optical filaments. Theoretical efforts to 
interpret the optical spectra have invoked various combinations of 
cooling or irradiation from the ICM, shocks, active galactic nucleus (AGN) 
emission or outflows, 
magnetic reconnection, and hot stars.  However, none of these 
scenarios has proven entirely satisfactory in explaining the 
observations; consequently, the origins and energetics of these 
systems remain poorly understood.

An important example of optical filaments in apparent association with 
a cooling flow is found in NGC~1275 (Kent \& Sargent 1979), 
the central elliptical galaxy in 
the Perseus cluster (Abell~426).  The nebular gas extends to at least 20 
kpc from the center of the galaxy (assuming H$_0$=75 km s$^{-1}$ 
Mpc$^{-1}$), with a complicated spatial structure; the optical 
spectrum fits within Heckman et al.'s Type II category.  The 
surrounding intracluster medium is believed to have a cooling flow, 
with mass deposition rates of up to 130 M$_\odot$ yr$^{-1}$ (Fabian, 
Cowie, \& Grindley 1981; Allen et al. 1999).  NGC~1275 features additional 
unusual characteristics 
that may be relevant to interpreting the extended nebular emission. 
The central elliptical galaxy is seen at a recession velocity of $\sim 
5200$ km s$^{-1}$, but is partially obscured by a smaller {\sl 
foreground} galaxy at a velocity of $\sim 8200$ km s$^{-1}$ (Minkowski 
1957); the redshift difference is consistent with the velocity 
dispersion observed for this cluster (e.g., Kent \& Sargent 1983).
NGC~1275 is also host to a strong AGN and associated radio source, 
3C~84, which emits jets that may be interacting with the ambient gas 
(Pedlar, Booler, \& Davies 1983; Pedlar et al. 1990).

In this paper, we report the results of a detailed study of the 
physical properties of the optical filaments associated with NGC~1275. 
This work makes use of extensive spectroscopic observations to quantify 
two-dimensional aspects of the observed emission, and also places sensitive 
limits on weak spectral features of diagnostic interest.  The results 
can be used to place stringent limits on the role of various agents 
proposed for heating or ionization of this plasma. 

\section{\sc Observations and Data Reduction} 
Long-slit spectra of the filaments of NGC~1275 were acquired in 10
observing runs during 1988 -- 1990, using the Shane 3-m telescope + UV
Schmidt CCD spectrograph at Lick Observatory (Miller \& Stone
1987).  The resulting data cover the wavelength range 3396 --
6940~\AA\ in three bandpasses, spanning 3396 -- 5000~\AA, 4695 --
6280~\AA, and 6130 -- 6940~\AA.  The slit was 2.5${\arcsec}$ wide and
132${\arcsec}$ long, with spatial sampling of 0.66${\arcsec}$~pixel$^{-1}$.
Observations were acquired with the slit centered on the nucleus and
oriented at four different position angles (PA = 62\arcdeg,
77\arcdeg, 283\arcdeg, and 313\arcdeg) selected to lie
along filaments with the highest surface brightness in H$\alpha$ (see
Figure 1).  Typical atmospheric seeing was 2\arcsec and not all
observations were done under strictly photometric conditions. The
spectral resolution was $\sim$3.5 \AA\space longward of 6130 \AA\space
and $\sim$7 \AA\space elsewhere.  The giant \ion{H}{2} region detected
by Shields \& Filippenko (1990a) lies at PA = 62\arcdeg. Line
emission from the high-velocity system ($cz \approx 8200$ km s$^{-1}$)
appears primarily northwest of the nucleus (e.g., Unger et al. 1990); 
it is visible in our spectra acquired at PA = 313\arcdeg, and to a
lesser extent at PA = 283\arcdeg.

The data were reduced using standard techniques, as implemented in IRAF
\footnote{The Image Reduction and Analysis Facility (IRAF) is distributed by 
the National Optical Astronomy Observatories, which is operated by the
Association of Universities for Research in Astronomy, Inc. (AURA),
under cooperative agreement with the National Science Foundation.}.
Spectra acquired at the same position angle and wavelength setting
were averaged after flux and wavelength calibration and sky
subtraction.  Estimates of the sky contribution were derived from
regions near the ends of the slit, where no nebular emission was
evident and the contribution of galaxy starlight was minimal.
Measurements of filament emission were obtained for regions where
H$\alpha$ emission was detected in a visual inspection of the 2-dimensional 
(2-D) spectra.

Emission-line strengths for the stronger lines were measured along the slit 
on a pixel-by-pixel basis, using an automated algorithm.  For each
resulting 1-D spectrum, the continuum in each wavelength setting was
approximated by a low-order polynomial fit to emission-free
bandpasses.  After subtraction of the continuum, candidate emission
lines were measured by fitting Gaussian profiles to all features
detected at $\ge 3\sigma$ significance.  The lines were subsequently
identified, and spurious features rejected, by comparison of the
measured list with an input list of line wavelengths, shifted by an a
priori estimate of the emission redshift; this redshift value was
measured interactively for H$\alpha$ in the spectrum of the filament
coadded along the slit.  Lines were identified if the predicted and
observed wavelengths agreed to within 5 \AA.  This bound was set to
allow for the correct identification of lines in the presence of
substantial ($\le 300$ km s$^{-1}$) velocity gradients that are
observed in this gas (e.g., Rubin et al. 1978).  The line flux
measured by this method is $\sqrt{2\pi}\sigma_{\lambda}F_p$, where
$\sigma_{\lambda}$ is the line profile dispersion as determined by the
Gaussian fit and $F_p$ is the peak height of the emission line above
the continuum. 

Estimation of line-intensity ratios for the NGC~1275 nebulosity requires
correction for reddening.  The filaments are subject to a Galactic
foreground extinction of $A_V = 0.82$ mag (Lockman \& Savage 1995) as
well as internal extinction, some of which may be due to the high-velocity 
system.  The limited signal-to-noise ratio (S/N) of many of the
pixel-by-pixel spectral extractions translates into significant
uncertainties in a reddening value based on Balmer-line ratios, and as
a result we derived an average reddening correction for each filament
(i.e., region of nebular emission seen on one side of the nucleus at a
given PA).  We obtained total extinction values from the measured
H$\alpha$/H$\beta$ line-intensity ratio, assuming an intrinsic value
of 2.86 (Hummer \& Storey 1987) and the extinction curve of Cardelli, Clayton, 
\& Mathis  (1989).  The results were compared with extinction values
implied by line ratios employing H$\gamma$ and H$\delta$, and were
found to be consistent, with the exception of measurements for PA =
313\arcdeg, where the limited S/N made the high-order 
Balmer lines unmeasurable, and for PA = 283\arcdeg. The line ratios
at the latter PA show evidence of underlying stellar absorption, which
is not surprising in light of earlier reports of early-type stellar
contributions to the integrated light seen in some portions of
NGC~1275 (e.g., Wirth et al. 1983).  The influence of the stellar
features is apparently strongest at this PA since the emission
equivalent widths are relatively weak. For this filament we determined
a consistent reddening value for the higher Balmer transitions subject
to absorption with constant equivalent width.  The total extinctions
(including the Galactic contribution, and neglecting the redshift
difference of the absorbers) calculated by these means are $A_V =$
1.0$\pm$0.2, 0.8$\pm$0.2, 2.3$\pm$0.5, and 1.1$\pm$0.4 mag for 
filaments at PA = 62\arcdeg, 77\arcdeg, 283\arcdeg, and 313\arcdeg, 
respectively.  The larger value of extinction towards PA = 283\arcdeg\
may be caused by the high-velocity system (Unger et al. 1990).  The
B-band atmospheric feature appears slightly longward of redshifted
[\ion{S}{2}]$\lambda$6731, but does not affect measurements of that
line.

In addition to studying the spatially resolved emission at each PA, we 
generated a composite spectrum of the filament system in order to
study weak emission features.  For each PA, the 2-D spectra were
averaged over intervals of high S/N filament emission (S/N
$\gtrsim$ 5 for H$\alpha$).  For PA = 62\arcdeg, this criterion
translated into an extracted region 14${\arcsec}$ wide, centered
23${\arcsec}$ away from the nucleus.  The PA = 77\arcdeg~and
PA = 283\arcdeg~extractions were 18.5${\arcsec}$ wide, centered at
21${\arcsec}$ and 20${\arcsec}$ away from the nucleus, respectively.
The extraction for PA = 313\arcdeg~spanned 33${\arcsec}$, centered
at 26${\arcsec}$ from the nucleus; we ultimately decided not to
include this extraction in the composite spectrum because the
emission-line equivalent widths were generally less than for the other
PAs, with a correspondingly lower S/N. The
resulting 1-D spectra were shifted to the rest wavelength
scale, as determined from the redshift of the Balmer lines, and the
spectra for different PAs were then combined to a single overall average. The 
relative velocity shifts between the filament extractions were less than 
90 km s$^{-1}$. The final composite spectrum (Figure 2) represents a total 
integration time of 10 hours in each wavelength setting.

In order to improve the accuracy of the continuum subtraction for the
composite spectrum, we experimented with modeling the stellar
continuum with templates derived from observations of emission-free
galaxies.  A reasonable correspondence in terms of the strength of
stellar features in NGC~1275 was found with the spectrum of NGC~205.
The stellar velocity dispersion of NGC~1275 is 250 km s$^{-1}$ (Nelson
\& Whittle 1995) while that of NGC~205 is 50 km s$^{-1}$ (Held, Mould,
\& de Zeeuw 1990).  We consequently convolved the NGC~205 spectrum with
a Gaussian having $\sigma = 245$ km s$^{-1}$ in order to match the
dispersion seen in NGC~1275.  To optimize the template match,
a fit with $\chi^2$ minimization was carried out using SPECFIT as
implemented in IRAF (Kriss 1994), varying the template redshift,
scaling, and reddening.  A best fit was obtained with $A_V = 0.6 \pm
0.1$ mag.  The Galactic extinction for NGC~205 is $A_V = 0.10$ mag
(Burstein \& Heiles 1984); if the two galaxies were perfectly matched
in spectral properties, this would imply a total extinction for
NGC~1275 of $A_V = 0.7 \pm 0.1$ mag, in agreement with the Galactic
foreground estimate.  This value is somewhat less than is obtained from 
Balmer emission lines for the filaments; the disparity may stem from 
real differences in the reddening of starlight and nebular
emission, and/or differences in stellar populations for NGC~1275 and
NGC~205.  The spectral regions over which we performed the continuum
fitting do not appear to be affected strongly by metallicity
differences between NGC~1275 and NGC~205.

For the stronger emission features, line strengths measured after
subtraction of this continuum fit were consistent with the results
obtained from a smooth fit to the continuum, as employed for the
pixel-by-pixel extractions.  Use of the galaxy template was important
for obtaining accurate values or limits for the weaker features,
however, and for some lines the template fit was locally optimized (by
adjusting the scaling) in order to match the observed continuum.  The
line fluxes obtained after subtraction of the continuum template are
listed in Table 1, and were obtained from Gaussian profile fits as
described above.  Error bars listed for the detected lines represent
the root-mean-square  uncertainty per pixel ($\sigma_c$) in the local 
continuum,
summed in quadrature over a width of 6$\sigma_\lambda$,
i.e., $\sqrt{6}\sigma_c\sigma_\lambda$. For lines that were not
detected, we list $3\sigma$ upper limits of
$\sqrt{2\pi}\sigma_\lambda$(3$\sigma_c$), where $\sigma_\lambda$ is
the width of either H$\beta$ or H$\alpha$, depending on which of these
Balmer lines is closer in wavelength.  For the composite spectrum, the
H$\alpha$/H$\beta$ ratio implies a total extinction of $A_V = 1.21$
mag, yielding an intrinsic value of $A_V=0.39$ mag.  The higher-order
Balmer lines in the continuum-subtracted spectrum showed evidence for
being over-corrected for the underlying stellar absorption: 
(H$\delta$/H$\beta$)$_{Observed}$ $>$ (H$\delta$/H$\beta$)$_{Case
B}$.  The implied disparity between the NGC~1275 and template spectra
does not significantly influence the estimate of $A_V$ obtained from
H$\alpha$/H$\beta$, however, due to the large equivalent widths of
these emission features.

\section{\sc General Results}

\subsection{\sc Line-Ratio Diagrams}

An informative overview of the emission properties of the NGC~1275
filaments can be obtained from standard line-ratio diagrams employed
for nebular classification (Baldwin, Phillips, \& Terlevich 1981;
Veilleux \& Osterbrock 1987).  Figure 3 shows three such
diagrams along with measurements from the pixel-by-pixel extracted
spectra for NGC~1275.  The observations have been corrected for
reddening, as described in \S2, and the use of intensity ratios of lines close
in wavelength makes these diagrams insensitive to any residual
reddening uncertainties.  Loci commonly adopted for classification of
\ion{H}{2} regions, Seyfert nuclei, and LINERs are also indicated
(e.g., Shields \& Filippenko 1990b).

As found in previous work, the filaments in NGC~1275 are characterized
by low-ionization emission, resembling that of LINERs (e.g., Heckman
et al. 1989).  An additional diagram illustrating the ionization state
of the gas is shown in Figure 4, which is a plot of
[\ion{O}{1}]$\lambda$6300/[\ion{O}{3}]$\lambda$5007 versus
[\ion{O}{2}]$\lambda$3727/[\ion{O}{3}]$\lambda$5007; this combination
is more sensitive to reddening, but has the benefit of tracing the
ionization of a single element.  Figure 4 shows that the filaments
overlap the low-ionization extreme of LINER behavior, and in
particular extend to very high [\ion{O}{1}]/[\ion{O}{3}] ratios.  We
note further that in Figure 3 the measurements do not
uniformly populate the LINER locus, but instead are clearly
concentrated to relatively low values for each of the
forbidden-line/recombination-line ratios.  Another way of stating 
this result is that the nebulosity is not merely described by low
ionization, as for LINERs, but is also characterized by low
excitation. Since the forbidden lines arise from collisional excitation, 
while the Balmer lines trace recombination, this result implicates 
a relatively low degree of heating per ionization in the cluster 
filaments. 

An interesting result illustrated by the 2-D spectra is that
the distribution of points shows a larger variation in the
[\ion{O}{3}]/H$\beta$ ratio than in ratios employing the forbidden
lines of lower ionization, such as [\ion{N}{2}]/H$\alpha$,
[\ion{S}{2}]/H$\alpha$, and [\ion{O}{1}]/H$\alpha$.  The measurements
include data for PA = 62\arcdeg~at the location of a giant
\ion{H}{2} region reported by Shields \& Filippenko (1990a).  The
ratios for this locale are distinct from those elsewhere in the
filaments, and contribute excursions to low values of
[\ion{O}{1}]/H$\alpha$ and [\ion{S}{2}]/H$\alpha$ seen in Figures 3$b$
and 3$c$.  Removal of these points would increase the contrast between
variation in [\ion{O}{3}]/H$\beta$ and in the other ratios.

\subsection{\sc Coronal Emission and Cooling}

A set of weak lines that are of special interest are the coronal
lines. These high-ionization collisionally excited lines are emitted from 
gas at a temperature of $\sim 10^{5.5-6.3}$ K, which is
intermediate between that of the ICM, 10$^{6-7}$ K, and the optical
filaments, $\sim 10^4$ K.  The coronal lines trace gas in the process
of cooling and their strength is a direct measure of the mass cooling
rate, $\dot{M}$ (Hu, Cowie, \& Wang 1985; Sarazin \& Graney 1991). Detection of
coronal lines may thus provide important diagnostics of the cooling-flow 
phenomenon.

The wavelength range covered in our spectra includes several coronal
lines, specifically [\ion{Ni}{12}]$\lambda$4232,
[\ion{Ar}{14}]$\lambda$4414, [\ion{Fe}{14}]$\lambda$5303,
[\ion{Ca}{15}]$\lambda$5445, [\ion{Fe}{10}]$\lambda$6374, and
[\ion{Ni}{12}]$\lambda$6702.  None of these transitions was detected
in our composite spectrum.  Upper limits to these lines were measured
after subtracting a fit to the stellar continuum using a galaxy
template of NGC~205, as described in \S 2.  The location of
[\ion{Fe}{10}]$\lambda$6374 coincides with the red wing of
[\ion{O}{1}]$\lambda$6364 and consequently warranted special
treatment.  The following procedure was used to remove the oxygen
line. A stellar continuum was fitted over the wavelength range of
6250 to 6350 \AA\, as described in \S 2, simultaneously with a 
Gaussian to [\ion{O}{1}]$\lambda$6300. To remove [\ion{O}{1}]$\lambda$6364, 
we divided the template of [\ion{O}{1}]$\lambda$6300 by the line ratio of
3.0 dictated by atomic parameters (Osterbrock 1989),
and subtracted the resulting profile and continuum fit from the composite 
spectrum.  An upper limit for
[\ion{Fe}{10}]$\lambda$6374 was derived from the residual. The upper
limits on the coronal lines were corrected for extinction corresponding
to  $A_V=1.2$ mag, the same as that applied to the detected lines.  The 
results are listed in Table 2.

Utilizing theoretical models and our upper limits, we can calculate
upper limits on the cooling rate of the hot ICM as measured in our
apertures.  The coronal line expected to have the highest emissivity
is [\ion{Fe}{10}] $\lambda$6374, for which Sarazin \& Graney (1991)
predict $1.39\times10^{37}$~erg~s$^{-1}$ for every 1
M$_\odot$~yr$^{-1}$.  For the one-pixel equivalent aperture
corresponding to the composite spectrum, the upper limit to the
reddening-corrected surface brightness of this line that we measure is
$\sim 2.7\times10^{-17}$~ erg~s$^{-1}$~cm$^{-2}$~ arcsec$^{-2}$.  The
composite spectrum is averaged over regions that extend out to a
radius of $\sim$ 28${\arcsec}$.  Assuming that the coronal emission is
distributed uniformly within this radius, our measurement suggests 
an upper limit for the luminosity of [\ion{Fe}{10}] of $\sim
4.0\times 10^{40}$ erg s$^{-1}$, corresponding to a mass cooling rate
of less than 2900 M$_\odot$ yr$^{-1}$.  This value of $\dot M$ is well
above the accretion rate derived from X-ray observations; upper limits
to the other coronal features would lead to even less restrictive 
limits on $\dot M$.  These results underscore the practical difficulty
in using the coronal lines to constrain $\dot M$, even with the investment
of substantial amounts of observing time.

The strength of coronal emisson relative to H$\alpha$ is of interest
if the nebular filaments are energetically coupled to the cooling ICM.
A cold filament irradiated by an isobarically cooling gas reprocesses
the ionizing radiation into H$\alpha$ with an efficiency $\sim$3\%
(Voit \& Donahue 1990; Donahue \& Voit 1991). 
The H$\alpha$ strength would thus be 0.03
(5/2) $(kT/\mu m_p) \dot{M}$, where $m_p$ is the proton mass, $\mu$ is
the mean mass per particle in units of $m_p$, $T$ is the temperature
of the hot gas, and $k$ is the Boltzmann constant.  In this picture,
both the H$\alpha$ and coronal-line luminosities scale with $\dot M$,
leading to well-defined predictions for their ratios, which we list
in Table 3 (Voit, Donahue, \& Slavin 1994; Sarazin \& Graney 1991).

The observed upper limits to the coronal-line strengths relative to
H$\alpha$ are larger than those of the predictions, thus leaving open
the possibility that the filaments are ionized by the cooling ICM, a
scenario we examine in more detail in Section 4.  The same data set as
employed here was used by Shields \& Filippenko (1992) to obtain a
more restrictive limit on [\ion{Fe}{10}]$\lambda$6374, after binning
the data on the basis of H$\alpha$ surface brightness.  The use of
an intermediate surface brightness bin produced a somewhat higher S/N
ratio in the continuum.  Their upper
limit to [\ion{Fe}{10}]$\lambda$6374/H$\alpha$ was $1.4 \times
10^{-3}$, adjusted to use the same upper-limit definition and
reddening correction employed here.  This value is $\sim 36$\% less
than the theoretical prediction, which may implicate other ionization
sources for the filaments.  However, this conclusion assumes that the
cooling plasma irradiating the nebular gas lies within our measurement
aperture, rather than on larger scales, and that the coronal feature
displays the same velocity width as H$\alpha$. 
Similar observational upper limits for coronal lines from cooling-flow 
filaments in other clusters have been reported by Donahue \& Stocke (1994) 
and Yan \& Cohen (1995). 

An additional complication to the use of the coronal lines as diagnostics
of cooling stems from persistent uncertainties in the relevant atomic
data (e.g., Mohan, Hibbert, \& Kingston 1994; Pelan \& Berrington 1995; 
Storey, Mason, \& Saraph 1996; Ferguson, Korista, \& Ferland 1997; Oliva 
1997). As summarized by Voit et 
al. (1994) for [\ion{Fe}{14}]$\lambda$5303, alternative choices of
atomic parameter values can lead to order-of-magnitude differences in
the predicted emissivities for a cooling ICM.  As noted in Table 3,
our upper limit to [\ion{Fe}{14}]$\lambda$5303/H$\alpha$ is less than
the value predicted by Voit et al. (1994) but larger than that of
Sarazin \& Graney (1991).  An unambiguous interpretation of coronal-line 
limits or detections will require further improvement in the
atomic data.

\subsection{\sc Ionization Mechanisms} 

A question central to the understanding of cooling-flow filaments is
the issue of what ionizes and powers these systems.  A comparison of
the recombination expected for mass deposition rates obtained from
X-ray observations with the observed H$\alpha$ line luminosities makes
it clear that simple cooling of the gas with a transition through a
nebular phase is insufficient to explain the energy requirements of
the filaments in NGC~1275 and other objects (e.g., Heckman et
al. 1989). A variety of mechanisms have been proposed for the ionization of 
cooling-flow filaments. Below we compare the predictions of 
these scenarios with the results of our observational study, taking into 
account various physical constraints. 

\subsubsection{\sc Ionization by an AGN}

NGC~1275 shows evidence for a Seyfert-like nucleus (Seyfert 1943), as
is clear from the emission-line spectrum (e.g., Filippenko \& 
Sargent 1985) and nonstellar continuum from
the central regions.  If the extended nebulosity is photoionized by
this continuum, the ionization state of the gas will reflect the
ionization parameter, which we define here as the ratio of ionizing
photon and electron densities at the face of the irradiated cloud.
The line ratio [\ion{S}{2}]$\lambda$6717/[\ion{S}{2}]$\lambda$6731 is
sensitive to the electron density, but remains consistent with the
low-density limit ($n_e \lesssim 10^2$ cm$^{-3}$; Osterbrock 1989)
throughout the filaments.  While the density may vary below this
threshold within the measured regions, pressure arguments suggest that
large fluctuations are unlikely (e.g., Johnstone \& Fabian 1988;
Heckman et al. 1989).  If the nebular filaments are ionized by the
AGN, we would then expect that the ionization parameter would largely
vary in the same way as the ionizing flux, proportional to $r^{-2}$. The line
ratio that is the best diagnostic of the ionization state is
[\ion{O}{3}]$\lambda$5007/H$\beta$, since this ratio involves a
high-ionization line (e.g., Ferland \& Netzer 1983; Ho et al. 1993).
The filaments range from $\sim 10{\arcsec}$ to $\sim 30{\arcsec}$ in
projected distance from the nucleus.  If these structures lie in the
plane of the sky, then the change in the ionization parameter due to
geometric dilution of flux would be approximately one order of
magnitude, translating into a significant variation in
[\ion{O}{3}]/H$\beta$. Values of this line ratio as a function of radius
for each of the filaments are shown in Figure 5.  While negative
gradients are seen in some places, the variations at large radial
distance tend to be small and/or non-monotonic, a result that is 
in agreement with previous observations having more limited PA coverage
(Johnstone \& Fabian 1988).

The observed behavior may still be consistent with photoionization by
a central source if the physical distance from the nucleus is not
indicated by the projected distance.  Johnstone \& Fabian (1988)
constructed a detailed model to study the effects of geometry in the
filaments of NGC~1275, and were able to reproduce the observed line
ratios as a function of projected distance, but suggested some
disagreement with the observed surface brightness of the filaments.
The present data show reasonable consistency between measured
surface brightnesses and predictions for ionization by the nucleus.
The flux in the ultraviolet continuum measured by 
the {\sl International Ultraviolet Explorer} (Kinney et al. 1991) 
and X-ray measurements from {\sl Einstein}
(Branduardi-Raymont et al. 1981) imply a 2-point spectral index
between 2500 \AA\ (after dereddening) and 2 keV of $\alpha_{ox} =
-1.4$ (assuming $f_\nu \propto \nu^\alpha$).  Extrapolating the UV flux
assuming a power law with this slope implies an ionizing luminosity of
$2\times10^{54}$ photons s$^{-1}$.  Clouds with a density of 
$n_e = 100$ cm$^{-3}$ at a
characteristic distance of $\sim 7$ kpc will be described by an
ionization parameter of $U \approx 10^{-4}$, consistent with the
observed line ratios.  These numbers also predict for
ionization-bounded clouds an emission measure (EM) of $\sim 500$ cm$^{-6}$
pc, in reasonable agreement (given the uncertainties in this
calculation) with the observed (reddening-corrected) surface
brightness of H$\alpha$ of $2 \times 10^{-15}$ ergs s$^{-1}$ cm$^{-2}$
arcsec$^{-2}$, corresponding to EM $\approx 900$
cm$^{-6}$ pc.  However, despite this nominal agreement, ionization
by the AGN appears unlikely based on other line diagnostics that are
independent of projection.

A line of particular interest for this purpose is
\ion{He}{2}~$\lambda$4686 since it is a gauge of the hardness of the 
ionizing continuum.  Figure 6 shows the composite filament spectrum
for NGC~1275 in the region of \ion{He}{2} $\lambda$4686 together with
the template continuum. The observed 3$\sigma$ upper limit for
\ion{He}{2} $\lambda$4686/H$\beta$ is 0.024, after dereddening. Photoionization
simulations show that this value is sensitive to the ionization
parameter for a given spectral shape.  Based on the continuum and
nebular parameters described above, we examined this issue using
photoionization calculations obtained with CLOUDY, version 90.04
(Ferland 1996).  The result is a prediction that
\ion{He}{2} $\lambda$4686/H$\beta \approx 0.10$, which is four times our 
upper limit. Therefore, the nebular filaments in NGC~1275 are highly 
unlikely to be powered by the AGN. 

In addition to being a source of ionizing radiation, the AGN in
NGC~1275 is a radio source (3C~84) with extended radio structures
(Pedlar 1981 et al.; Pedlar et al. 1990), that may in principle power
the optical nebulosity.  The relation between the radio plasma and the
nebular gas was discussed briefly by McNamara, O'Connell, \& Sarazin
(1996) in a multiwavelength study dealing with NGC~1275 and its
environs. The extended radio emission roughly coincides with two
cavities in the X-ray gas located on the northeast and southwest outer
portions of the galaxy (Pedlar et al. 1990; McNamara et al. 1996). The
nebular emission seems to avoid the radio emission, which may be
evidence for the radio lobes displacing the nebular gas. Another
interpretation would be that the nebular gas, which has a higher
density, may be confining the radio-emitting gas and channeling it to
regions of lower density (McNamara et al. 1996). In any case, if an
interaction with the radio lobes were responsible for powering the
nebular emission, we might expect the optical filaments to be correlated
spatially with the radio structures.  This correspondence is not seen,
but might indicate that the shocked gas remains in a hot phase.
Recent observations with {\sl Chandra}, however, show little evidence
for X-ray emission by a hot shocked component in the regions in question
(M. Wise, private communication); a similar situation has been
reported for Hydra A (McNamara et al. 2000).  As will be argued below,
these results and additional spectroscopic constraints suggest that
shock heating is unlikely in the filaments of NGC~1275.

\subsubsection{\sc Shocks}
The above arguments point toward a distributed ionizing source as the
underlying power source for the filaments in NGC~1275. This ionization 
mechanism could in principle be shocks, stars, and/or the cooling ICM
itself. Several observational properties can be interpreted as possible
evidence of shocks in the filaments.  The velocity field of the nebular gas
shows variations of $100-300$ km s$^{-1}$ (e.g., Burbidge \& Burbidge 1965; 
Heckman et al. 1989). Moreover, the line widths are greater than the thermal
width. Line emission from gas at a temperature of 10,000 K will exhibit
a full width at half maximum (FWHM) of $\sim 20$ km s$^{-1}$ 
for hydrogen, which can be compared with observed FWHM values of
$100-300$ km s$^{-1}$ in the spatially-resolved spectra, after removal of 
the instrumental width.

Several possible mechanisms are available for driving shocks,
including interaction of the radio plasma from the AGN with
surrounding gas (\S 3.3.1), pressure imbalances resulting from
thermally unstable cooling in the cluster center (Cowie, Fabian, \&
Nulsen 1980), and accretion infall onto the central galaxy.  For
shocks driven by radio ejecta, Bicknell, Dopita, \& O'Dea (1997) have
shown that the luminosity in H$\alpha$ is of order 0.1\% of the
energy flux into the radio plasma, with the exact value dependent on
the gas density distribution and shock velocity.  For NGC~1275, 
Pedlar et al. (1990) estimate that $\sim 10^{43}$ erg s$^{-1}$
is injected into the radio halo; this value is inversely related
to the age of the halo, which Pedlar et al. take to be $10^8$
years or less.  The total H$\alpha$ luminosity of the NGC~1275 filaments
is $\sim 2.1 \times 10^{42}$ erg s$^{-1}$ (Heckman et al.
1989; Caulet et al. 1992).  It appears doubtful that the radio
source is sufficient to power the observed line emission, unless
the radio halo is exceedingly young or the energy transfer to
optical emission is unusually efficient.

Efficient cooling of matter near the cluster center can lead to
regions that are underpressured relative to their surroundings, and
hence will be subjected to repressurizing shocks (Cowie, Fabian, \&
Nulsen 1980).  The power delivered by such shocks results from the
thermal and mechanical energy of the ICM, so that the maximum power
that be can extracted from the cooling ICM is (5/2) $(kT/\mu m_p) \dot{M}
\approx 10^{43}$ erg s$^{-1}$.  Only about 3\% of this energy is
processed into H$\alpha$, requiring that the cooling clouds be shocked
multiple times to reproduce the observed emission (the ``$H_{rec}$
problem''; see Heckman et al. 1989 for a discussion).  The apparent
survival of grains in the filaments of other cooling-flow galaxies
suggests that passage of repeated shocks through the nebular gas is
unlikely (Donahue \& Voit 1993).

If the bulk kinetic energy of infalling matter dominates the energy
inflow to the central region, the power available for driving shocks
is of order $\dot{M} v_i^2$, where $v_i$ is the characteristic infall
speed.  If $v_i$ is as high as the cluster velocity dispersion ($\sim
1000$ km s$^{-1}$; Kent \& Sargent 1983), the resulting rate of energy
transfer is $\sim 10^{44}$ erg s$^{-1}$, which is sufficient to power
the filament emission.  However, if $v_i$ is closer to the filament
velocity dispersion, the available power is an order of magnitude less,
and probably insufficient to generate the observed H$\alpha$ luminosity.

One general characteristic of shocks is that line-intensity ratios of emitted
radiation reflect the kinetic energy flux entering the shock front,
and hence are a function of shock velocity (e.g., Shull \& McKee 1979;
Dopita \& Sutherland 1995, Binette, Dopita, \& Tuohy 1985).  If the
observed line width reflects the characteristic shock velocity, then
we would expect the line ratios to correlate with line width; such
behavior is in fact seen in large-scale outflows from starburst
galaxies (e.g., Heckman et al. 1999).  A plot of
log([\ion{N}{2}]$\lambda$6583/H$\alpha$) versus the FWHM of H$\alpha$ from 
the four PAs is shown in Figure 7, together with model predictions,
which exhibit a positive trend in most cases. The range of observed FWHM
is $100-300$ km s$^{-1}$, while [\ion{N}{2}]/H$\alpha$ is clustered
around 0.9. A shallow but statistically significant correlation
(Spearman non-parametric significance of $\sim 5\times 10^{-3}$) is
present, with larger line ratios associated with higher velocity
width.  This trend is due to combining values from all the PAs; 
PA = 283\arcdeg~and PA = 313\arcdeg~have slightly higher line ratios and 
FWHM than PA = 62\arcdeg~and PA = 77\arcdeg. Significant correlations are 
not seen in either the individual
PAs or in PA62/PA77 and PA283/PA313 groupings, making shocks as
heating sources in the filaments unlikely.  Similar behavior is seen
in the [\ion{O}{3}]/H$\beta$ ratios, although no correlation or
PA-dependent behavior is evident in [\ion{S}{2}]/H$\alpha$ or
[\ion{O}{1}]/H$\alpha$.

A further piece of evidence which argues against shocks comes from the
upper limit on electron temperature as determined from the
[\ion{O}{3}] lines (Table 1).  Figure 8 shows the continuum around the
expected location of [\ion{O}{3}]$\lambda$4363.  An upper limit on the
temperature as measured from
[\ion{O}{3}]$\lambda$4363/[\ion{O}{3}]$\lambda$$\lambda$4959, 5007
$<$0.047 restricts the electron temperature to be less than 33,000 K
(Osterbrock 1989; Shaw \& Dufour 1995).  Shock models predict
[\ion{O}{3}]$\lambda$4363/[\ion{O}{3}]$\lambda$$\lambda$4959, 5007 to
be between 0.01 and 0.07 (Shull \& McKee 1979; Raymond 1979; 
Binette et al. 1985;
Dopita \& Sutherland 1995), corresponding to electron temperatures
between 13,000 K and 54,000 K. Shock models with velocities less than
$\sim 300$ km s$^{-1}$ predict an [\ion{O}{3}] line ratio consistent
with our upper limit, but have problems in simultaneously matching the 
high- and low-ionization forbidden-line strengths. These models also 
consistently underpredict \ion{He}{1} $\lambda$5876, except at high 
velocities, where the predicted 
\ion{He}{2} $\lambda$4686 exceeds our upper limit (Raymond 1979; 
Binette et al. 1985; Dopita \& Sutherland 1995). Therefore, we conclude that 
shock heating in the filaments of NGC~1275 is a minor contributor to
the energetics of the nebular gas.

\subsubsection{\sc Stellar Photoionization}

Cooling-flow galaxies commonly exhibit an optical continuum that is
bluer than typical elliptical galaxies (e.g., Johnstone, Fabian, \&
Nulsen 1987; McNamara \& O'Connell 1989; Allen 1995; Cardiel, Gorgas,
\& Aragon-Salamanca 1995, 1998).  Some degree of correlation exists
between such blue continua and nebular emission, prompting suggestions
that both may trace the influence of an underlying young stellar
population.  While nebular emission (bremsstrahlung, free-bound, and
2-photon) may contribute to such a continuum, a quantitative
calculation indicates that the continuum emission accompanying the
observed H$\beta$ luminosity would be 2 orders of magnitude too weak
to account for the observed light.  Stars are thus the likely source
of the blue continua.  As possible ionizing agents for the nebular
gas, stars also have the added appeal of producing negligible
\ion{He}{2} nebular emission, consistent with the observations of
NGC~1275 and other cooling-flow galaxies.

Early studies of the large-scale nebular morphology of NGC~1275 
interpreted this object and its system of radial filaments as the
result of a large-scale outflow; indeed, it was called 
an ``exploding galaxy'' (Burbidge \& Burbidge 1965).
Modern studies of another object given this label, M82, demonstrate
that it is in a sense ``exploding'' because of its 
large-scale outflow driven by star formation (e.g., McCarthy et al. 1987; 
Shopbell \& Bland-Hawthorn 1998). However, such vigorous star formation 
is not apparent in NGC~1275.

NGC~1275 does show weaker indications of star formation within the
last $\sim 10^7 - 10^8$ yrs on a variety of scales (Wirth et al. 1983;
Holtzman et al. 1992; McNamara et al. 1996).  Optical measures of the
young population have been used to estimate roughly the associated
ionizing luminosity, yielding a result that is consistent with the
requirements of powering the filaments (Wirth et al. 1983; Romanishin
1987).  However, there is not a strong spatial correspondence between
the regions of recent star formation and the nebular gas.  As noted
previously, the nebular spectrum also differs markedly from typical
\ion{H}{2} region spectra; in the one instance where recent star
formation is clearly evident within a filament (at PA = 62\arcdeg),
the nebular line ratios are distinct from the remainder of the
filament system (Shields \& Filippenko 1990a).  Photoionization models
suggest that hot stars can generate LINER-like emission (Filippenko \&
Terlevich 1992; Shields 1992), but these scenarios generally have
difficulty producing [\ion{O}{1}] and [\ion{N}{1}] emission as strong
as seen here, unless the gas contains a very high density component
($\sim 10^5$ cm$^{-3}$).  

Stars may nonetheless play a role in powering the filaments if they
generate a radiation field that ionizes gas on scales exceeding that
of localized \ion{H}{2} regions.  A similar scenario has recently
gained favor for explaining the ionization of the diffuse ionized gas
(DIG) commonly seen in spiral galaxies (e.g., Howk \& Sembach 1999;
Rand 1998; Martin 1997).  However, the forbidden-line strengths seen
in the DIG are characteristically weaker than in the NGC~1275
filaments; typical DIG line ratios are
[\ion{O}{1}]$\lambda$6300/H$\alpha \lesssim 0.03$,
[\ion{N}{2}]$\lambda$6583/H$\alpha
\approx 0.5$, \ion{He}{1} $\lambda$5876/H$\alpha$ $\lesssim 0.02$, and 
[\ion{O}{3}]$\lambda$5007/H$\alpha \lesssim 0.1$ (Galarza, Walterbos,
\& Braun 1999; Sembach et al. 2000), which contrasts markedly with the
values in Table 1. Another aspect which makes the cooling-flow
filaments of NGC~1275 distinct from the DIG is the emission measure.
For the DIG, EM $\approx 1-10$ cm$^{-6}$ pc (Reynolds 1985; Martin \&
Kennicutt 1997; Hoopes, Walterbos, \& Rand 1999), and can reach values
as high as 100 cm$^{-6}$ pc (Galarza et al. 1999; Hoopes et al. 1999).
In contrast, the average EM in NGC~1275 for the regions studied here
is $\sim$ 900 cm$^{-6}$ pc.

Given the arguments for and against stellar photoionization, 
it may be the case that this excitation process, if present, is at work in 
combination with another ionization source, as has been pointed 
out by Voit \& Donahue (1997) for Abell~2597. The evidence from NGC~1275, 
however, does not form a strong case for stellar photoionization as the 
dominant heating mechanism in the filaments. 

\subsubsection{\sc Photoionization by the ICM}

The hot ICM surrounding NGC~1275 can be the ionization source of the
filaments if its energy is channeled efficiently into the nebulosity
(Voit \& Donahue 1990; Begelman \& Fabian 1990).  
In one such scenario, known as self-irradiation, the hot ICM ionizes 
nebulae that have already cooled and condensed from the ICM itself (Donahue
\& Voit 1991; Voit \& Donahue 1990; Voit et al. 1994). 
Another way to channel ICM energy into the nebulae is the formation of
mixing layers between cold clouds ($T_e\approx 10^4$ K) embedded in a
hot ambient medium ($T_e\approx 10^7$ K) (Begelman \& Fabian 1990;
Crawford \& Fabian 1992). Line emission comes from the mixing layer,
which is described by an intermediate temperature
($T_e\approx 10^{5.5}$ K), and from the cool cloud that is now being
irradiated by the mixing layer.  Crawford \& Fabian (1992) were able
to reproduce the line ratios of Abell~496, a cluster that displays Type I
filaments, with a mixing-layer prescription. However, in order to
reproduce the line ratios of the Type II cluster Abell~2597, these authors
found it necessary to invoke an additional energetic contribution from
shocks.
 
The observed line ratios in NGC~1275 are close to values predicted by
one of the self-irradiation models presented by Voit et al. (1994).
In that model (D39T67), the ICM is cooling from a temperature of
5$\times$10$^6$ K, the ionization parameter is 10$^{-3.9}$, and the
column density of irradiated \ion{H}{1} is 10$^{21.5}$ cm$^{-2}$. 
Their predicted \ion{He}{2} $\lambda$4686/H$\beta$ ratio is almost a 
factor of 3 larger than our upper limit. This comparison shows that
self-irradiation is not the ionization mechanism because it provides a
continuum which is harder than that inferred from the observed line
ratios. A cooling-flow system that shows similar filaments, in terms
of line ratios, to NGC~1275 is Abell~2597 (Heckman et al. 1989; Voit
\& Donahue 1997).  Voit \& Donahue (1997) similarly ruled out a hard
ionizing continuum as the energy source in the filaments of Abell~2597
on the basis of an upper limit of 0.02 on \ion{He}{2}
$\lambda$4686/H$\beta$, which is again a factor of $\sim 3$ below
expectations for self-irradiation.

\subsubsection{\sc Composite Ionization Models}

None of the ionization mechanisms considered individually in the
preceding sections is entirely satisfactory in accounting for the
emission properties and energetics of the cooling-flow filaments in
NGC~1275. Given the compelling evidence for the presence of young
stars, it is reasonable to explore models that are composites of
stellar photoionization and some secondary heating mechanism. The
study by Voit \& Donahue (1997) of Abell~2597 is illuminating in this
regard. Voit \& Donahue (1997) invoked stellar photoionization as the
ionizing mechanism for the optical filaments of Abell~2597 after placing
strong upper limits on [\ion{O}{3}]$\lambda$4363 and
\ion{He}{2} $\lambda$4686, a result that ruled out shock heating and
self-irradiation by the ICM. However, they found that the electron
temperature predicted for stellar photoionization was lower than what
they inferred from their observations of the [\ion{O}{2}] and
[\ion{S}{2}] lines. For this reason, they suggested that a heating
source in addition to hot stars may be important.  The secondary
source they considered was the ICM; it can heat the filaments via
electron conduction (e.g., Sparks 1992), acoustic-wave heating (e.g., 
Pringle 1989), and MHD-wave heating (e.g., Friaca et al. 1997).  Voit
\& Donahue (1997) argued that the resulting energy flux incident on
the nebulae can become greater than that of photoelectric heating if
the nebulae are characterized by low ionization, a case which holds in
cooling-flow filaments. Unfortunately, detailed theoretical predictions
appropriate for comparison with observations have not yet been
generated for these scenarios.  A related case where the heating rates
due to the dissipation of turbulence have been calculated and found to
be comparable to the radiative cooling rates is in the DIG of the
Milky Way (Minter \& Spangler 1997). However, detailed observational
parameters (such as the amplitude of density fluctuations, and
turbulence length scales) needed to calculate the heating rate from
this mechanism are unavailable for cooling-flow filaments.
 
We investigated an alternative scenario in which the observed emission
arises from a combination of clouds subject to stellar photoionization
and irradiation by the ICM. In this simplified model, we assumed that
within a given measurement aperture, we see a mix of clouds ionized by
stars and by radiation from the ICM. In this picture we observe a
weighted average of the spectrum of the two types of cloud.  A
fraction of the observed line emission comes from regions ionized by
stars, while another fraction comes from nebulae irradiated by the
ICM. This weighted average translates into the corresponding fractions
in terms of line ratios (with all ratios normalized to H$\beta$). We
used line ratios from the self-irradiation models of Voit et
al. (1994), and representative values for intermediate-metallicity
\ion{H}{2} regions (Veilleux \& Osterbrock 1987). Once the relative
contributions of stars and the ICM were determined from
\ion{He}{2} $\lambda$4686/H$\beta$, attention was turned to reproducing
[\ion{O}{1}]$\lambda$6300/H$\alpha$ as well as the other line
ratios. [\ion{O}{1}]$\lambda$6300/H$\alpha$ is a good diagnostic
because it is very weak in \ion{H}{2} regions.  Including more stars
pushed [\ion{O}{1}]$\lambda$6300/H$\alpha$ below observed values,
while more self-irradiation would push
\ion{He}{2}~$\lambda$4686/H$\beta$ above what is observed. Ultimately
we were not able to find a composite model that fits all the line
ratios simultaneously. Combinations satisfying the helium requirement,
with marginal agreement for the other line ratios, failed to match the
oxygen constraint, and vice versa. Composite models can be found that
satisfy both [\ion{O}{1}]/H$\alpha$ and \ion{He}{2}/H$\beta$, but
these scenarios proved inconsistent with the other line ratios. The
reason for this failure to come up with a successful hybrid model may
be due to our simplistic approach, which neglects clouds that see
ionizing radiation from {\sl both} sources simultaneously, or more
likely suggests that other ionization mechanisms are at work.

\subsubsection{\sc Attenuated Cluster Radiation}

If the filaments in NGC~1275 and other cooling-flow galaxies
are photoionized, hints about the nature of the ionizing continuum come from 
the strengths of the emission lines. Relatively strong forbidden 
lines, the absence of \ion{He}{2}~$\lambda$4686, and the presence of 
\ion{He}{1}~$\lambda$5876 suggest that the ionizing continuum cuts 
off at energies somewhere between 24 and 54 eV. An ionizing source
generating a continuum with a natural cutoff at 54 eV is normal O-type
stars. For this reason we tried in the previous subsection 
to construct a composite model of stellar photoionization and 
self-irradiation by the ICM, but our attempts were unsuccessful in terms of 
matching the observed line ratios. 

The cooling ICM can produce a range of ionizing continua, depending on
its initial temperature. Self-irradiation models (Voit et al. 1994)
that predict a \ion{He}{2} $\lambda$4686 value similar to what is
observed fail at reproducing the other line ratios, while
self-irradiation models that agree with most of the observed line
ratios have high \ion{He}{2} $\lambda$4686/H$\beta$. One way to
construct an appropriate ionizing spectrum with the shape dictated by the
observations is to introduce a cutoff at the helium edge
of 54 eV by attenuating the spectrum by an ionized screen. The
filtering medium will imprint a sharp dip at 54 eV while leaving the flux
of photons with energies between 24 and 54 eV nearly unchanged. 
Observational evidence for such a screen may come from X-ray observations
of Abell~426 that suggest the presence of absorbing matter with a 
column density of
$N_H\approx 10^{21}$ cm$^{-2}$, in excess of the Galactic foreground
component; the ionization state of this medium is not well understood
(e.g., Allen, Fabian, \& Edge 1992; Arnaud \& Mushotzky 1998).

We therefore constructed models in which the ionizing continuum from
the cooling ICM is attenuated through a screen before irradiating the
filaments.  This scenario differs from the traditional
self-irradiation models of Voit \& Donahue (1990) in the sense that
the filaments and the screen need not be the by-products of the
cooling flow.  The models discussed here also differ from the mixing
layers picture of Crawford \& Fabian (1992) in terms of the origin of
the ionizing continuum. In the mixing layers model, the ionizing
spectrum comes from gas at a temperature intermediate between the
temperature of the ICM and that of the optical filaments.

To test the photoionization plus screen scenario, we carried out
photoionization calculations using CLOUDY, version 90.04 (Ferland
1996).  We assumed plane-parallel geometry throughout the
calculations, and a metallicity of 0.4 times solar values, consistent
with X-ray studies (e.g., Allen et al. 1992).  The flux originating from 
the ICM was scaled to match the soft X-ray central surface
brightness reported by Mohr, Mathiesen, \& Evrard (1999), and
propagated consistently through the calculation.  We found good
results assuming a temperature for the dominant ionizing plasma of
10$^6$ K.  While lower than that of the majority of the ICM, this
value is in general accord with the high degree of cooling occurring
in the Perseus cluster core, and with spectroscopic evidence for cooler
phases (e.g., Mushotzky \& Szymkowiak 1988) in this region.

In order to have the appropriate screening properties, the ionization
state of the filtering medium should be such that the optical depth at
54 eV is large enough to suppress \ion{He}{2} $\lambda$4686 in the
filaments, while the total hydrogen column density stays less than, or
comparable to, the column density inferred from X-ray
observations. Moreover, the ionization parameter of the screen should
be high enough to produce minimal optical forbidden-line emission,
since the screen is evidently not seen in such lines.  Therefore, we
chose the hydrogen density in the screen to be 1 cm$^{-3}$, implying
an ionization parameter of $U \approx 10^{-1.1}$, and stopped the
calculation when the optical depth at 54 eV became 10. The resulting
hydrogen column density in the screen is $\sim$6$\times$10$^{21}$
cm$^{-2}$.  For the filament, the forbidden-line spectrum suggests
an ionization parameter $U \approx 10^{-3.9}$ (\S 3.3.1).  This value
emerges naturally from the propagated radiation field and a choice of
density of $\sim 100$ cm$^{-3}$, consistent with the observed
[\ion{S}{2}]$\lambda$6717/$\lambda$6731 line ratio.

Resulting predictions for the emission-line spectrum emergent from a
filament are shown in Table 3. All the line ratios were satisfactorily
reproduced, with the exception of the [\ion{S}{2}] lines, which are
too strong by a factor of $\sim 1.6$. Figure 9 shows the continuum
incident on the screen, and the continuum exiting the screen and
incident on the filament. Notice the strong ionized helium edge at 54
eV. Since the ionized screen is photoionized, it will also emit
recombination radiation, especially in \ion{He}{2} $\lambda$4686
(\ion{He}{2} Pa$\alpha$).  In a sense, then, this model removes the
difficulty of excess \ion{He}{2} emission in the filaments, but
probably creates a new observational inconsistency in that the
screen should be a strong source of observable recombination emission,
which is not apparently seen.

\ion{He}{2}~$\lambda$4686 emission would be reduced by a factor of
$\sim 2$ if recombination in the screen were described by Case A
conditions rather than Case B.  However, the line center optical depth
for \ion{He}{2} Ly$\gamma$, which upon reabsorption can branch to emit
the Pa$\alpha$ line, remains $\gg 1$ for the column density and
temperature describing our model screen; additional turbulence of
$\sim 10^4 - 10^5$ km s$^{-1}$ within the screen would be required to
yield Case A conditions.  Case A recombination thus does not appear
to be a plausible mechanism for reducing emission in \ion{He}{2}~$\lambda$4686.

\subsubsection{\sc Diffuse Extreme Ultraviolet Radiation}

Observations with the {\sl Extreme Ultraviolet 
Explorer} \/({\sl EUVE}) indicate the presence of 
excess soft X-ray/EUV emission in clusters of galaxies (Lieu et
al. 1996; Lieu, Bonamente, \& Mittaz 1999). The existence and
properties of this emission are still highly debated (e.g., Arabadjis
\& Bregman 1999). If this emission component is real, an interesting
question is whether it plays a significant role in the energetics of
the nebular filaments.

A difficulty with using the EUV observations directly stems from the
fact that {\sl EUVE} data (Lieu et al. 1996) cover only a limited
bandpass, making any attempts at determining the total ionizing
luminosity dependent upon the shape of the assumed continuum, a
property that has to be determined by a model for the origin of the
EUV excess.  According to one interpretation, inverse-Compton
scattering of the cosmic microwave background radiation off cosmic-ray
electrons is responsible for the EUV continuum, which is then
described by a two-point spectral index between 70 and 912~\AA\ of
$\alpha_{EUV} = -2/3$ (Sarazin \& Lieu 1998; Sarazin 1999).  With this
spectral energy distribution, the ratio of photon surface brightness
between 70~\AA\ and 912~\AA\ to that between 70~\AA\ and 160~\AA, the
approximate range covered by the {\sl EUVE} DS Lexan detector, is
about 6.2.

No {\sl EUVE} observations have been made of NGC~1275; the large
Galactic \ion{H}{1} column density toward this source ($1.4 \times
10^{21}$ cm$^{-2}$; Lockman \& Savage 1995) makes its detection
difficult or impossible at EUV wavelengths.  In order to obtain an
order-of-magnitude estimate of the relevant photon flux in the {\sl
EUVE} bandpass, we used observations of Abell~1795 (Lieu et al. 1999).
Abell~1795 is a cooling-flow cluster whose central galaxy displays bright
optical filaments which, like those in NGC~1275, were classified
spectroscopically by Heckman et al. (1989) as Type II; the $\dot M$
values inferred from X-ray observations agree to within a factor of
$\sim 2$ for the two clusters.  Assuming an
effective area of $\sim 15$ cm$^{2}$ for {\sl EUVE} with the DS Lexan
detector (Miller 1997), the central surface brightness for Abell~1795 is
2.0$\times$10$^{-4}$ photons s$^{-1}$ cm$^{-2}$ arcmin$^{-2}$ (Lieu et
al. 1999).  Scaling this value to obtain the ionizing (70~\AA\ --
912~\AA) surface brightness results in a value of $\sim 1.2\times
10^{-3}$ photons s$^{-1}$ cm$^{-2}$ arcmin$^{-2}$.  If we assume that
the filaments can be treated as ionization-bounded slabs located at
the center of the EUV-emitting medium, the ionizing photon flux
striking the optical gas is $\sim 4.5\times 10^4$ photons s$^{-1}$
cm$^{-2}$, implying EM $\approx 0.06$ cm$^{-6}$ pc in the
photoionized slab. This value differs substantially from the emission
measure calculated above from our H$\alpha$ surface brightness for
NGC~1275, $\sim 900$ cm$^{-6}$ pc, which is also comparable to the
results obtained for other systems (Voit \& Donahue 1997).  We
conclude that the EUV emission, if present in Abell~426, is a highly
unlikely ionizing mechanism for the optical filaments in NGC~1275.

\subsubsection{\sc Magnetic Reconnection} 
Another source of energy in galaxy clusters is the intergalactic magnetic 
field. Magnetic reconnection of field lines in clusters can release 
energy in a manner similar to that of solar flares 
(e.g., Heckman et al. 1989; Soker \& Sarazin 1990).  
For magnetic reconnection to be effective in releasing 
energy, the magnetic field must be strong and tangled. Reconnection of 
magnetic flux loops will liberate energy which has to be included in the 
energy budget of the cooling-flow filaments (e.g. Soker \& Sarazin 1990; 
Jafelice \& Friaca 1996). 

Cooling flows present in galaxy clusters may amplify the magnetic
field energy density to equipartition values with the gas pressure
within the inner tens of kpc due to compression of the field lines
that are frozen-in to the plasma (e.g., Soker \& Sarazin 1990). It is
interesting to note that the observed optical filaments in cooling-flow 
galaxies occur over the same length scale predicted for
equipartition between magnetic and gas pressures, for theoretical
cooling-flow models.  The resulting strong magnetic fields are
expected to produce large Faraday rotations in the cores of
cooling-flow clusters, in agreement with radio measurements for many
sources (Ge \& Owen 1993, 1994).  As a result of this amplification,
one would expect clusters with larger cooling rates to have larger
Faraday rotation measures.

The correlation between cooling rates and Faraday rotation measure
seems to be borne out by observations.  Taylor, Barton, \& Ge (1994)
compiled a list of radio sources at the centers of galaxy clusters,
and showed that such a trend exists. Notable exceptions, however, are
NGC~1275 and PKS~0745-191, which have strong cooling flows and
extended optical nebulae, but no measurable polarization (Ge \&
Roberts 1993; Taylor et al. 1994).  One possible interpretation of
this finding is that a magnetized screen with complex structure is
strongly depolarizing the radio signal (Ge \& Roberts 1993; Taylor et
al. 1994). The screen may be the optical filaments themselves. Several
pieces of observational evidence indicate that this might be the
case. The H$\alpha$ and radio emission are cospatial in projection in
PKS~0745-191 and in the core of NGC~1275 (Heckman et al. 1989; Ge \&
Roberts 1993).  Van Breugel, Heckman, \& Miley (1984) noted that there
is a spatial inverse correlation between radio polarization and H$\alpha$
surface brightness in Abell~1795, a result later confirmed by Ge \& Owen
(1993) at higher spatial resolution. Obtaining more radio polarimetry
data with high spatial resolution for cooling-flow clusters will be
very helpful in testing this inverse correlation. A detailed rotation
measure map will make it possible to look for a relation between the
H$\alpha$ luminosity of a filament and the degree of depolarization it
imprints on the radio emission cospatial with it.

The above discussion provides circumstantial evidence for the
importance of magnetic fields in understanding cooling-flow
filaments. If Faraday depolarization in the optical filaments can be
taken as evidence for the filaments being magnetized with a complex
geometry, then magnetic field lines may be reconnecting in the nebular
gas, and one would expect strong depolarization to be associated with
strong optical emission.  In terms of energetics, Jafelice \& Friaca
(1996) argued, on the basis of theoretical models, that magnetic
reconnection is capable of powering the lower luminosity Type I
filaments, but falls short of sustaining the Type II systems, such as
NGC~1275, Abell~1795, and PKS~0745-191.  Additional high spatial
resolution radio polarimetry for cooling-flow galaxies is desirable in
order to carry out a comparative study of the magnetic properties of
the two source types.  A further interesting experiment would be to
obtain such data for the central galaxy in S\'{e}rsic 159-03 (S
1101), which displays filaments with emission-line ratios encompassing
both Type I and Type II values (Crawford \& Fabian 1992).

\section{\sc Conclusions}

In this paper we have presented the results of a detailed optical
spectroscopic study of the optical filaments around NGC~1275, the
central galaxy in the Perseus cluster of galaxies. The database was
extensive enough to yield tight upper limits on weak lines of interest
for constraining the ionization and thermal state of the nebular gas.
Line-ratio diagrams reveal that the nebulosity has an unusual
combination of emission-line strengths reflecting both low ionization
and low excitation.

The spectroscopic results were used to gauge the viability of 
several scenarios proposed for powering cooling-flow nebulosity.
Mechanisms examined include shocks, ionization by an AGN central
source or extended radio structures, stellar photoionization, 
and energy transfer by various means from the ICM.  None of these
models is entirely satisfactory in accounting for the energetic
and/or emission-line properties of the filaments.  Radio studies
of Faraday rotation for cooling-flow galaxies provide tantalizing
suggestions that magnetic reconnection plays a significant role in
the filament energetics.  Additional polarization studies of these
sources may provide a clearer picture as to the importance of
magnetic fields in these systems.

\acknowledgments
Support for this work was provided in part by a grant from the
California Space Institute to J.C.S., by NSF grants AST-8957063 
and AST-9417213 to A.V.F., and by NASA grant NAG5-3556 to A.V.F. 
We thank the Mount Hamilton staff of Lick Observatory for their
friendly and capable assistance over the multiple observing runs
required for this project. At the time when the observations were
obtained, Lick Observatory received partial funding from the NSF
through Core Block Grant AST-8614510 to the University of
California. We also thank Ray Bertram and Rick Pogge for assistance at
the Perkins 1.8-m telescope. We gratefully acknowledge Gary Ferland
for making CLOUDY available for this study, and Mike Wise for
communicating results from {\sl Chandra} observations in advance of
publication. 


\pagebreak
\begin{figure}
\vbox{
\centerline{
\psfig{figure=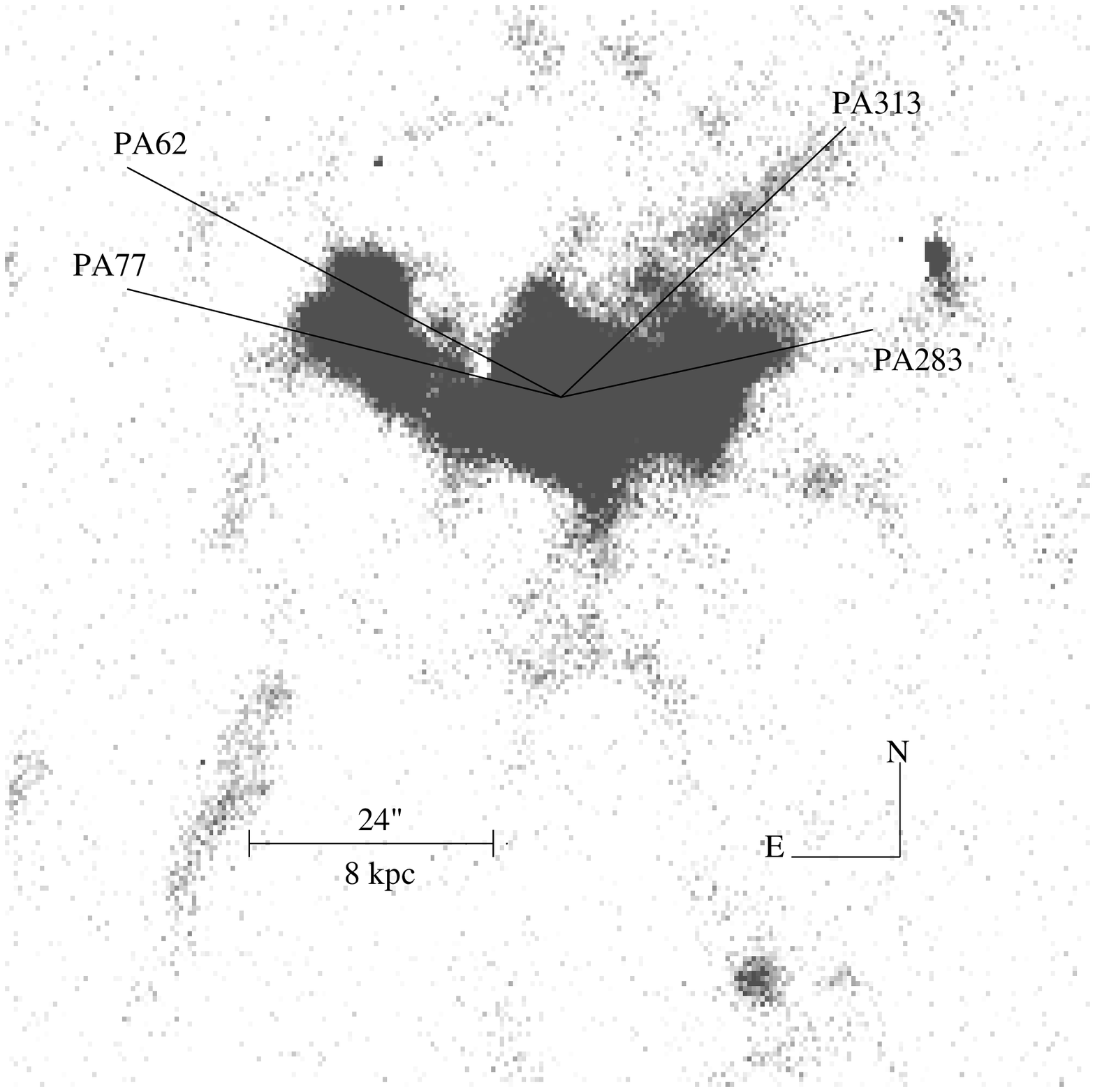,width=5.in}
}
\caption{
Continuum-subtracted H$\alpha$+[\ion{N}{2}] image of NGC~1275
obtained on 1992 January 2 U.T. by Shields and R. Pogge at the Perkins 
1.8-m telescope, showing 
the position angles of the slit employed for the measurements
presented here.
}\label{figone}
}
\end{figure}

\begin{figure}
\vbox{
\centerline{
\psfig{figure=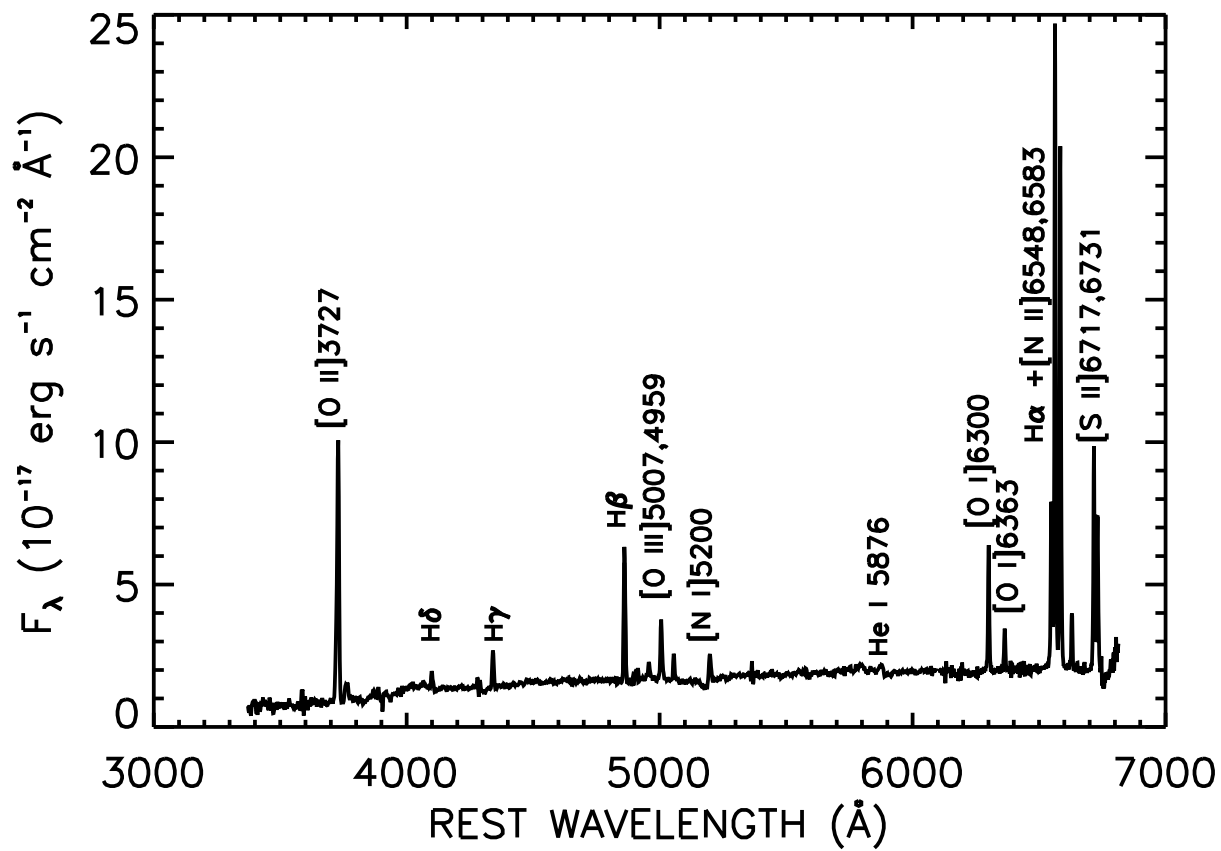,width=7.in}
}
\caption{
Composite spectrum generated from averaging filament spectra
obtained at PAs 62$^\circ$, 77$^\circ$, and 283$^\circ$. The emission 
features seen at 5057 \AA\ and 6629 \AA\ are 
[\ion{O}{3}]$\lambda$5007 and 
H$\alpha$, respectively, from the high-velocity system, redshifted by 
$cz \approx 3000$~km s$^{-1}$ relative to the low-velocity system.
}\label{figtwo}
}
\end{figure}
\begin{figure}
\vbox{
\centerline{
\psfig{figure=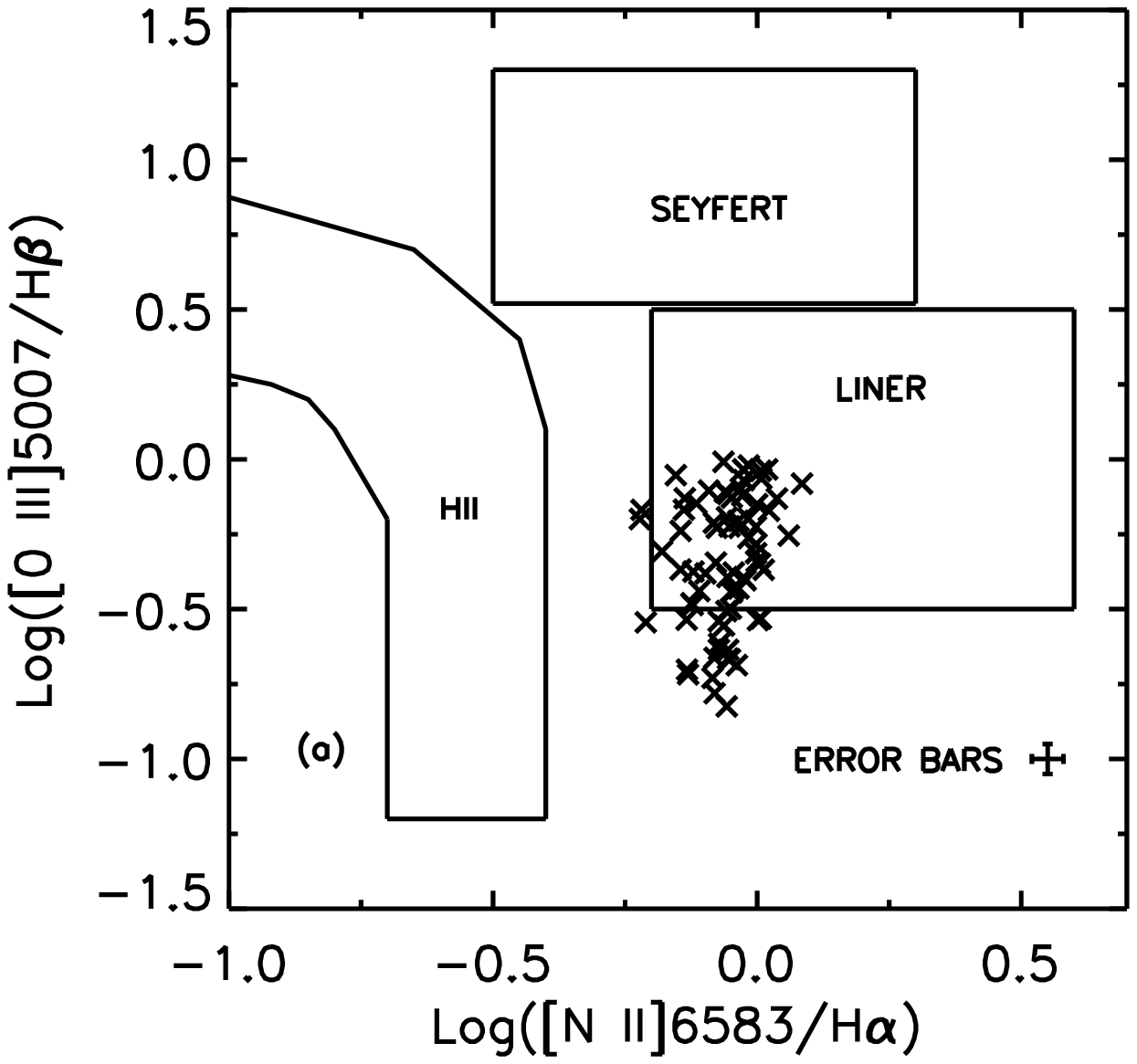,width=3.in}
\hspace*{0.5in}
\psfig{figure=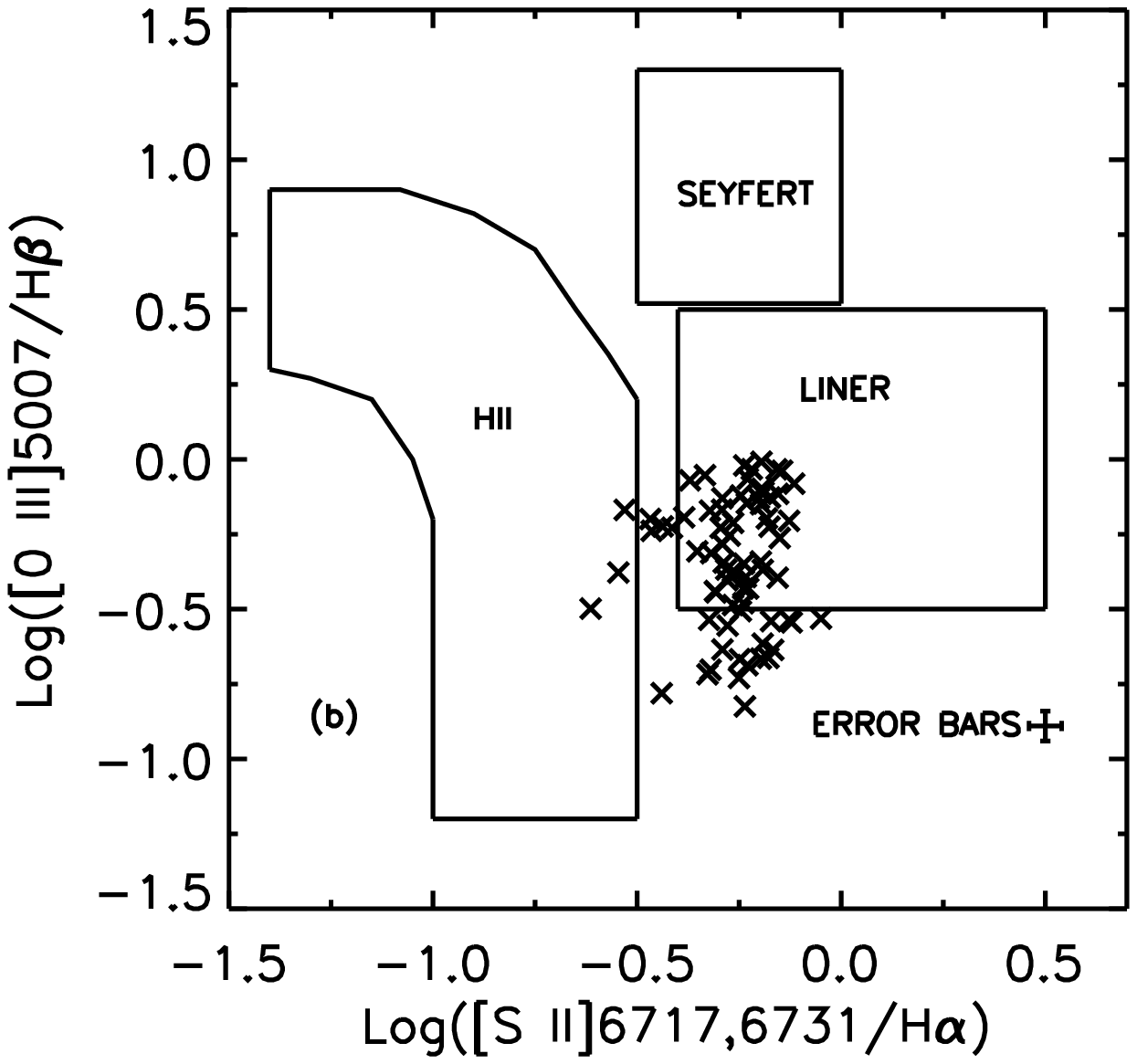,width=3.in}
}
\vspace*{0.5in}
\centerline{
\psfig{figure=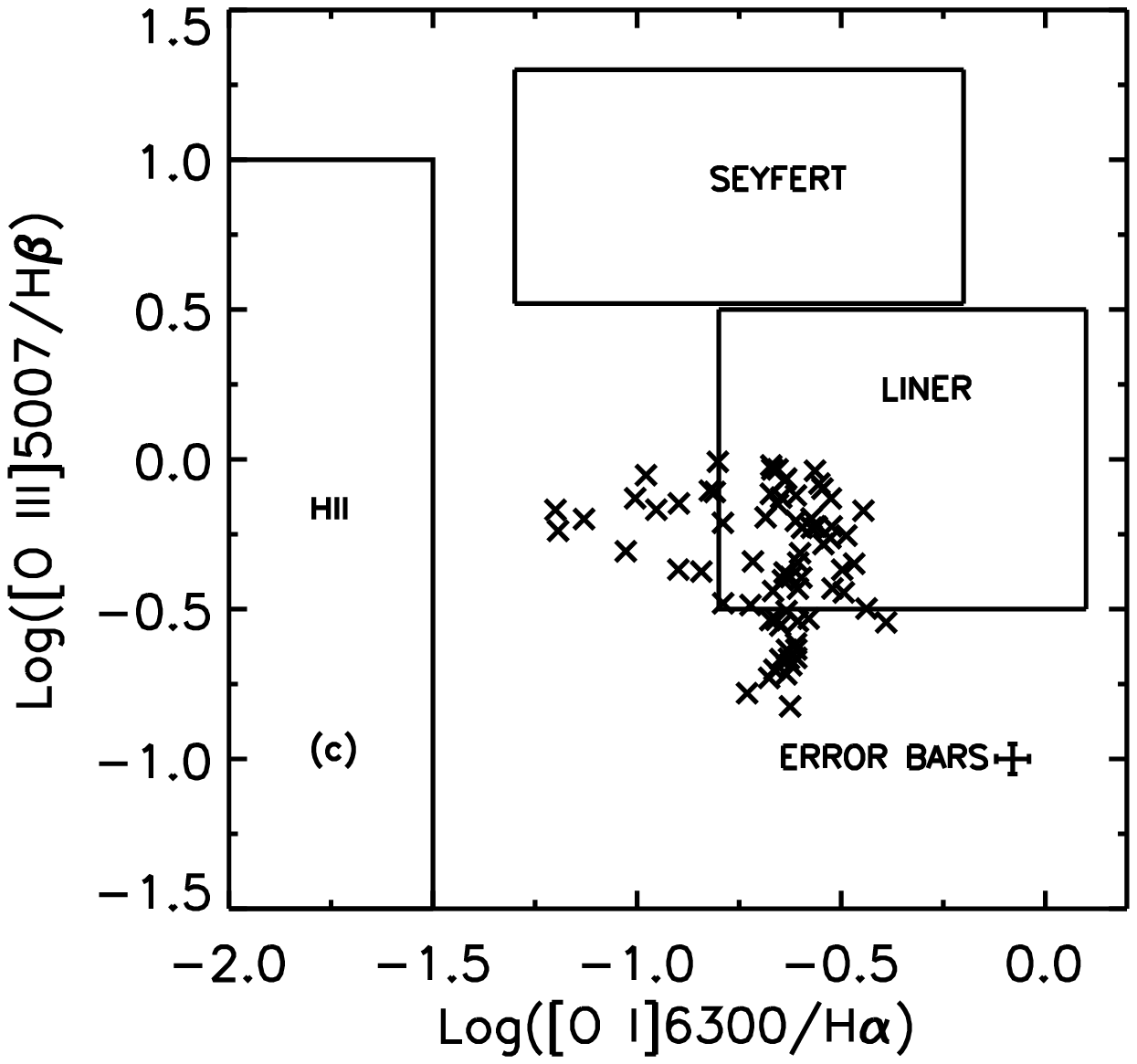,width=3.in}
\vspace*{0.3in}
}
\caption{
Line-ratio diagrams displaying reddening-corrected values measured at 
different locations within the filaments.  Approximate loci employed
for classification of nebular types are indicated.  Representative $1\sigma$
error bars are shown.
}
}
\label{figthree}
\end{figure}

\begin{figure}
\vbox{
\centerline{
\psfig{figure=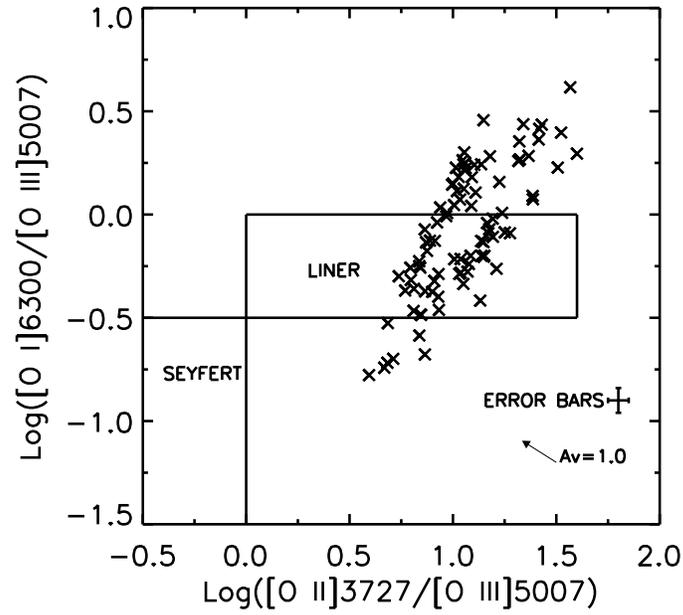,width=3.in}
\vspace*{0.3in}
}
\caption{
Line-ratio diagram employing lines of oxygen only, with reddening-corrected
filament measurements shown.  The effect of reddening by $A_V = 1$ mag
is given by the arrow.
}
}
\label{figfour}
\end{figure}

\begin{figure}
\vbox{
\centerline{
\psfig{figure=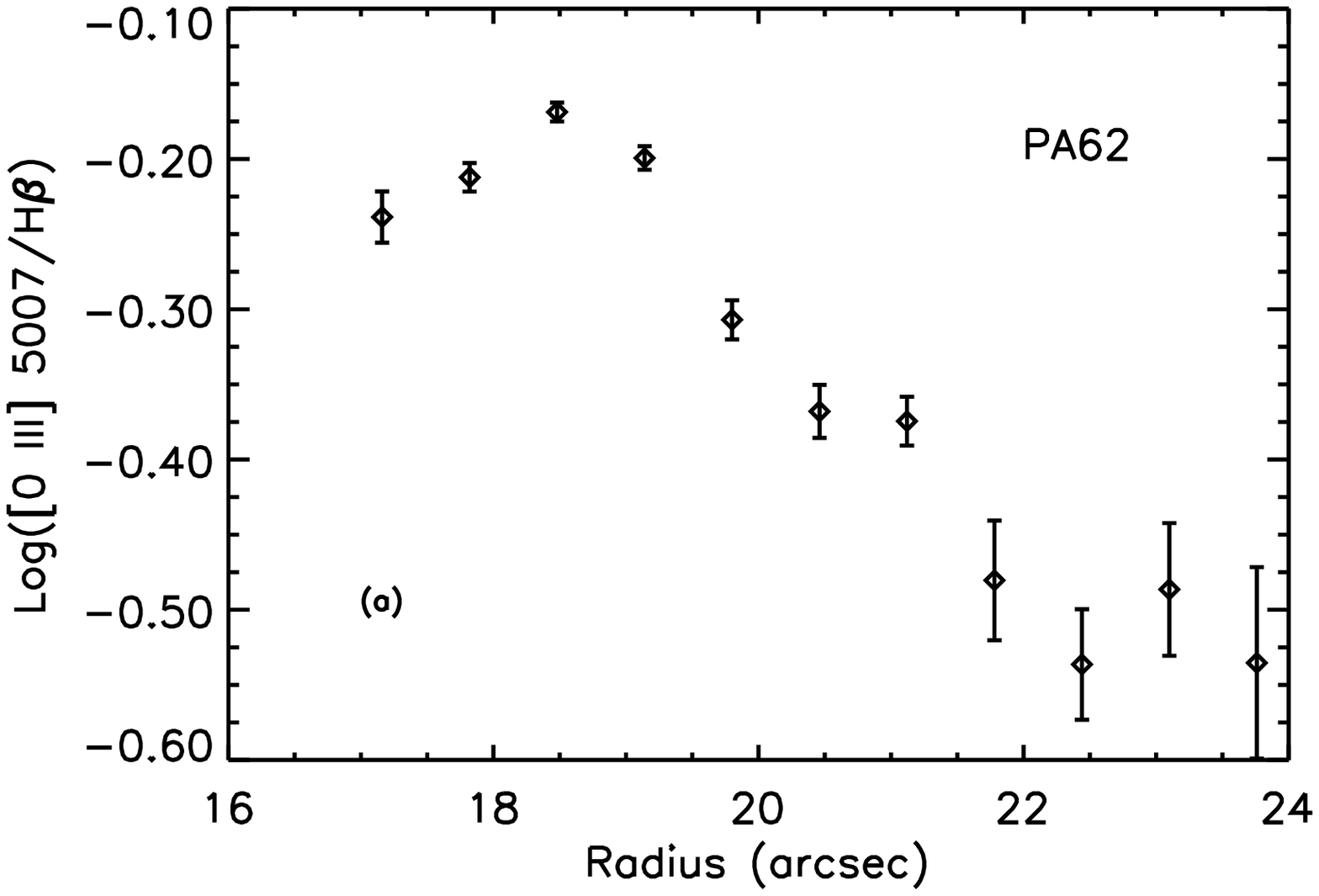,width=4.in}
\psfig{figure=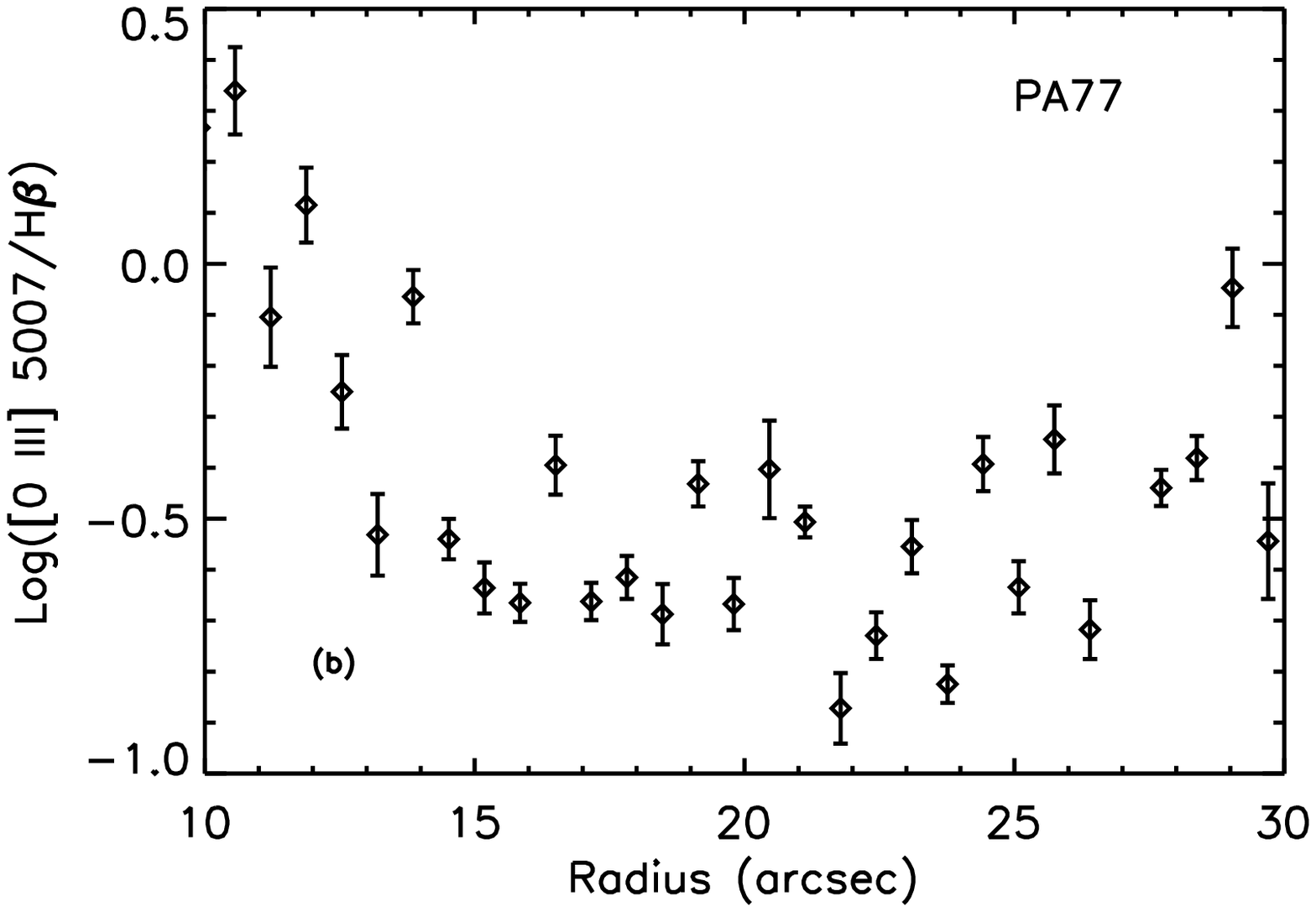,width=4.in}
}
\centerline{
\psfig{figure=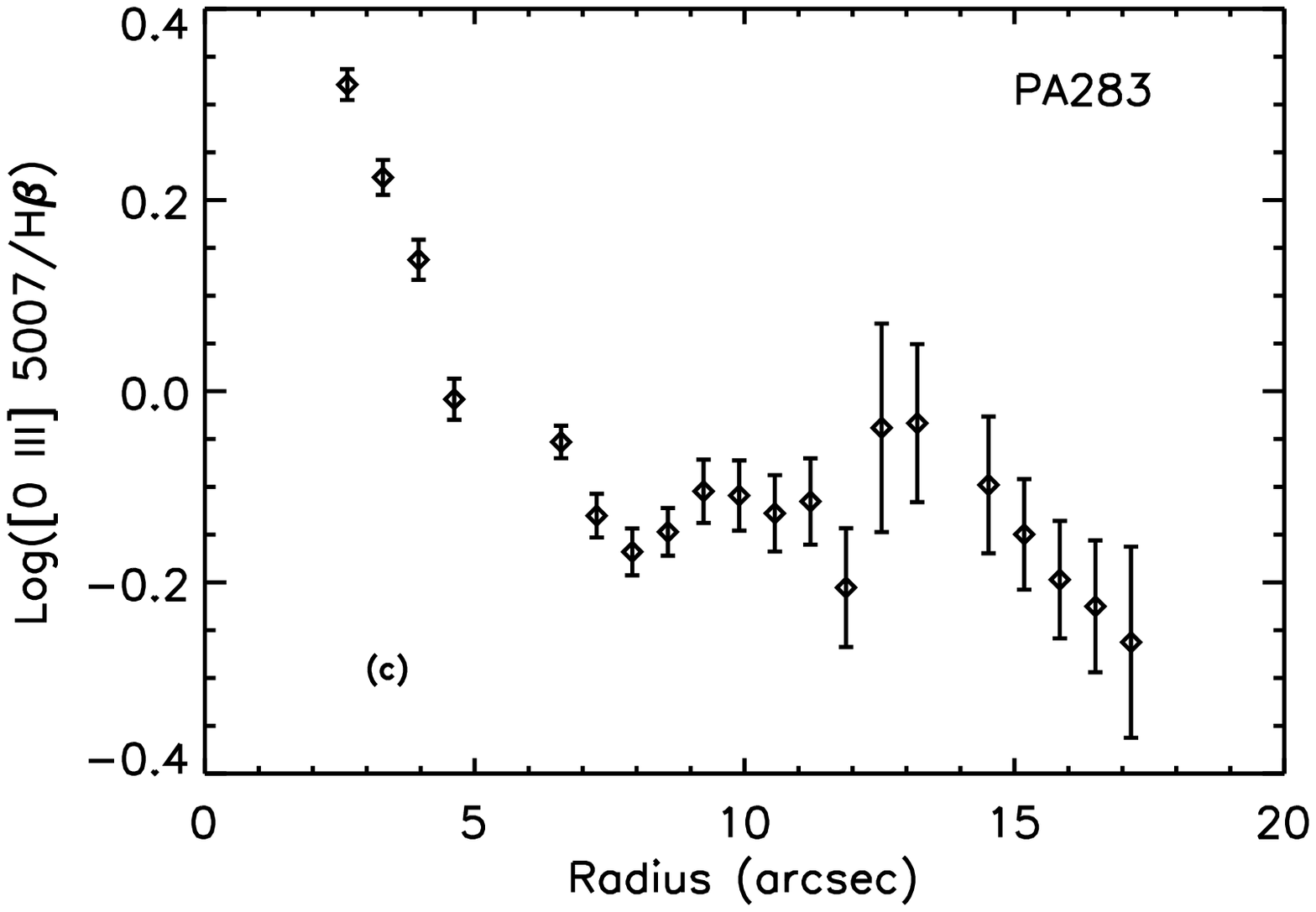,width=4.in}
\psfig{figure=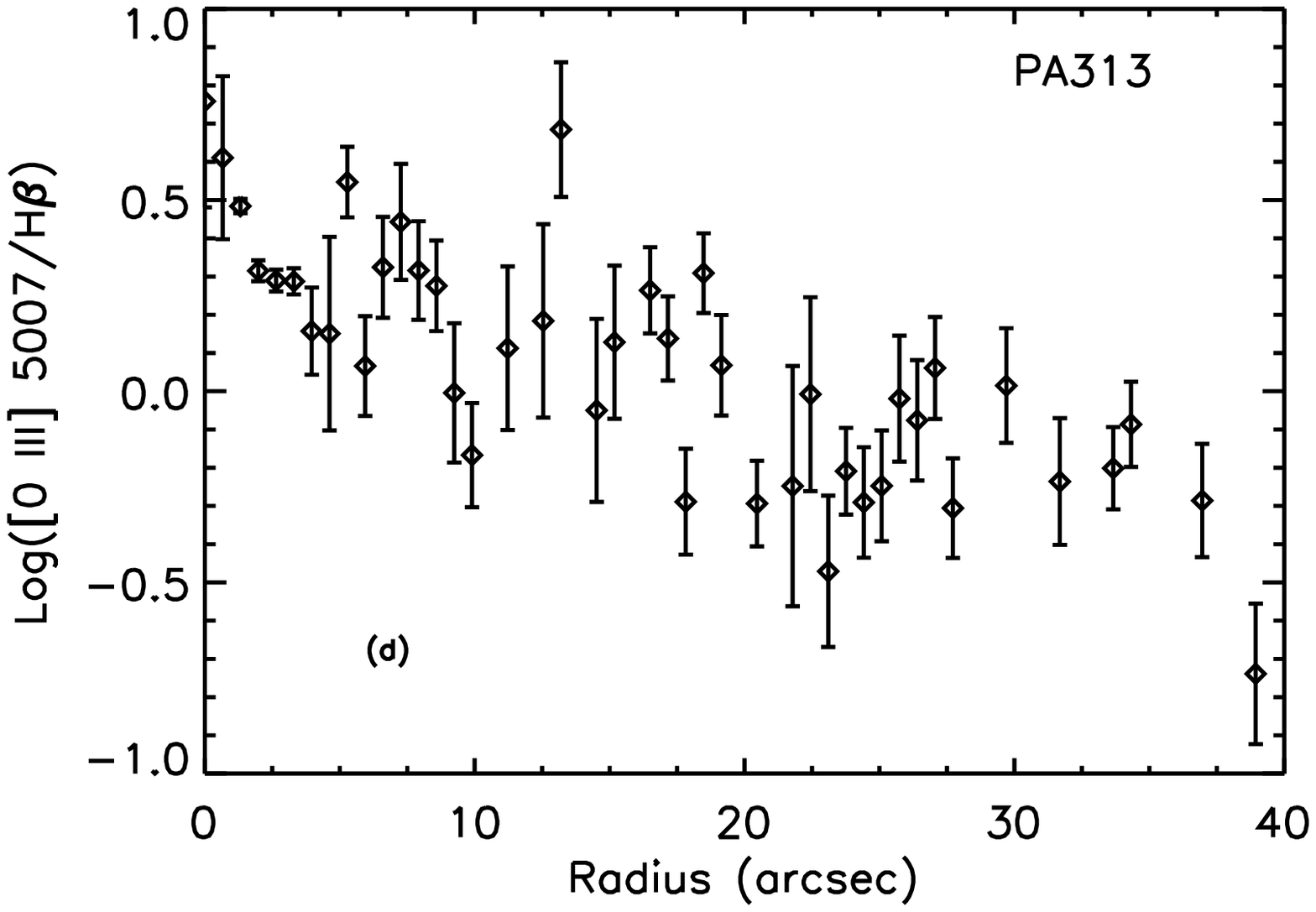,width=4.in}
}
\caption{
Radial profiles of the [\ion{O}{3}]/H$\beta$ line ratio at the four PAs;
error bars represent $1\sigma$ uncertainties.
}\label{figfive}
}
\end{figure}

\begin{figure}
\vbox{
\centerline{
\psfig{figure=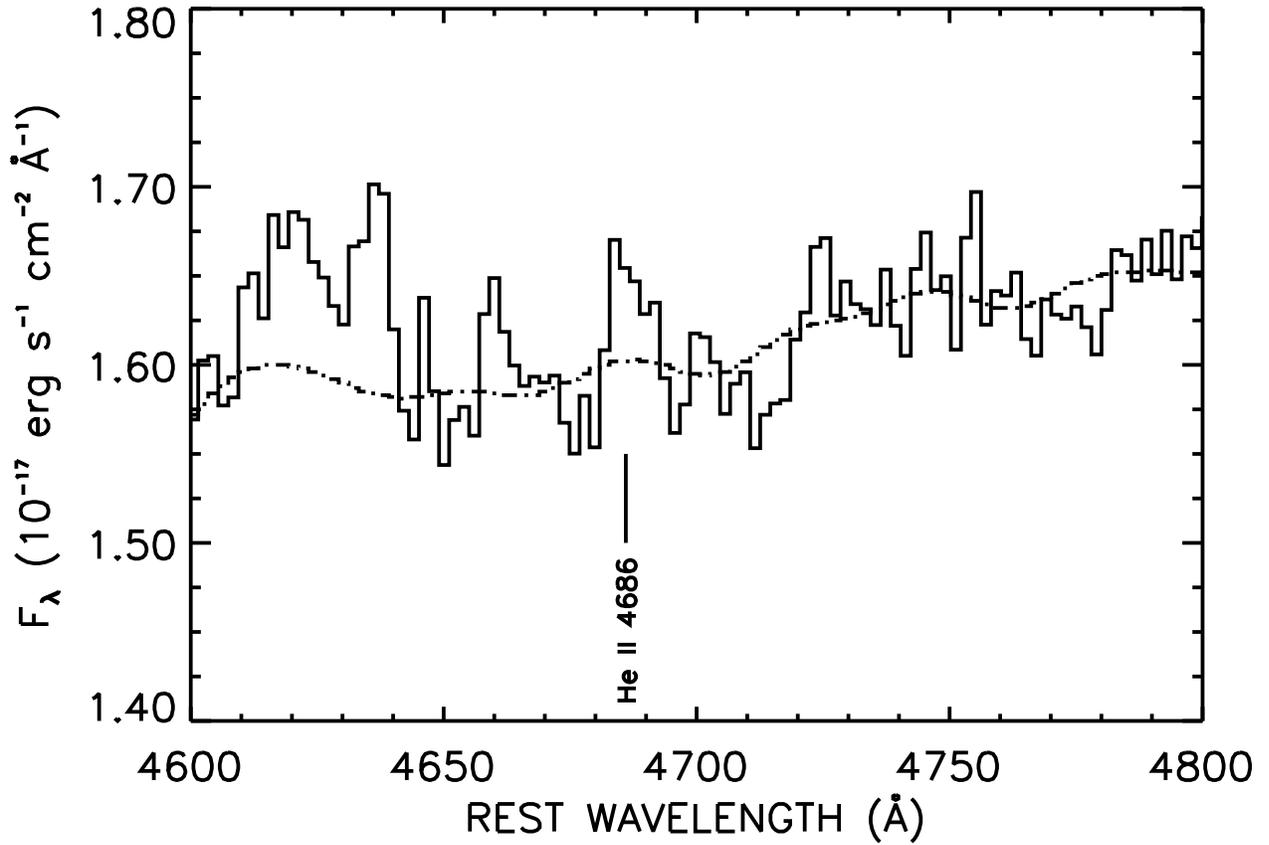,width=7.in}
}
\caption{
Expanded view of the composite spectrum from Figure 2, illustrating
the measurement for NGC~1275 (solid histogram) and best-fitting
continuum template (dashed line) in the vicinity of \ion{He}{2} 
$\lambda$4686. The expected location of this emission feature is shown.
}\label{figsix} 
}
\end{figure}
\begin{figure}
\vbox{
\centerline{
\psfig{figure=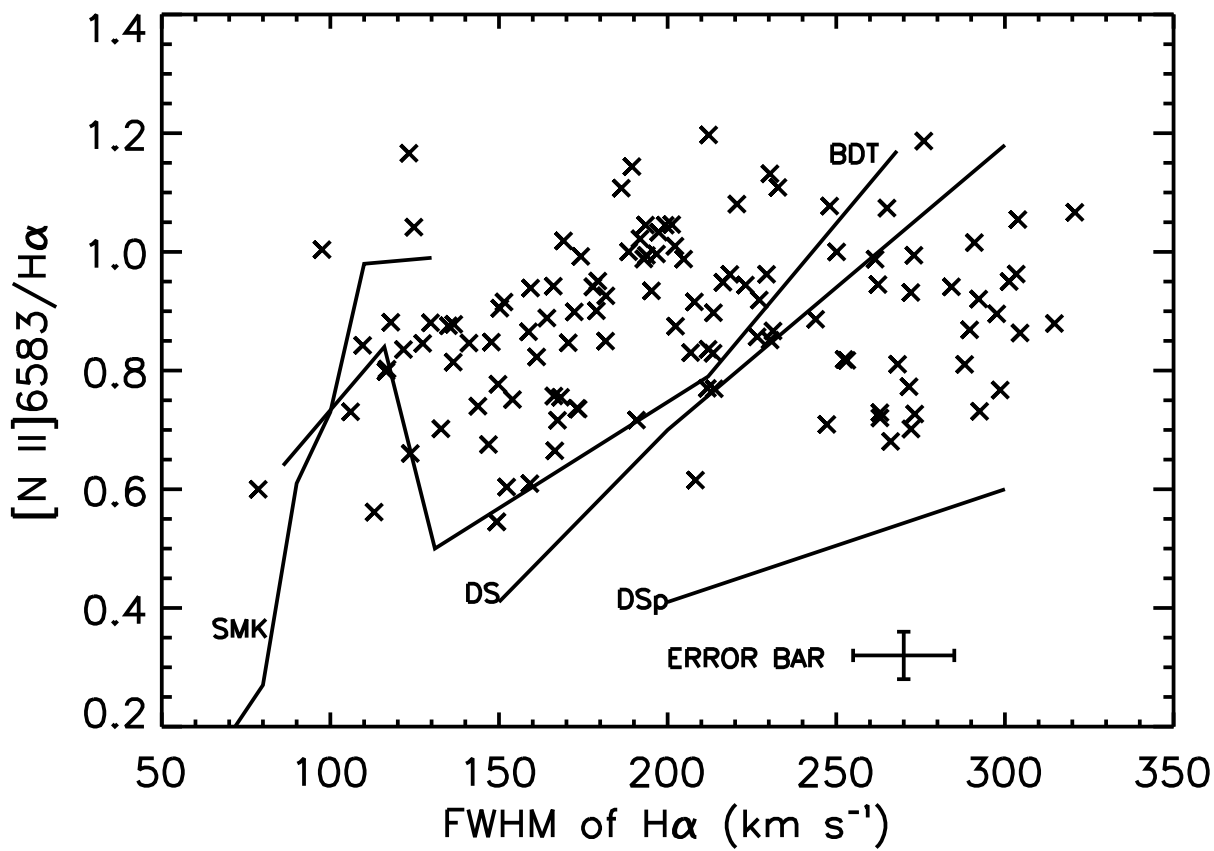,width=7.in}
}
\caption{
Reddening-corrected [\ion{N}{2}]$\lambda$6583/H$\alpha$ line ratio
as a function of H$\alpha$ line width for the filament measurements.
Also shown by the solid lines are predictions of shock models taken
from Dopita \& Sutherland (1995; DS = shock only, DSp = shock + precursor),
Binette et al. (1985; BDT), and Shull \& McKee (1979; SMK).
In plotting the model predictions, we have equated the shock velocity
to the FWHM of H$\alpha$.
}\label{figseven}
}
\end{figure}

\begin{figure}
\vbox{
\centerline{
\psfig{figure=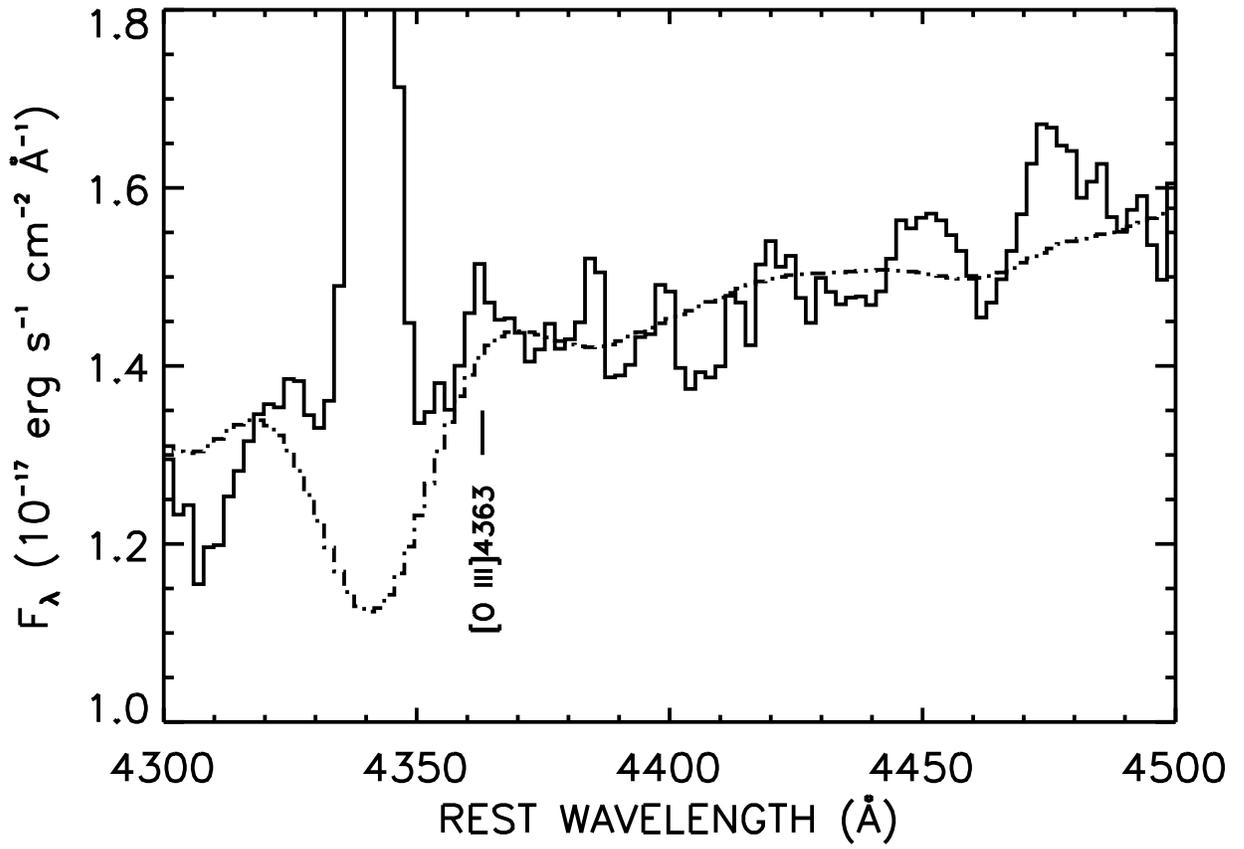,width=7.in}
}
\caption{
Similar to Figure 6, illustrating the spectral region around 
[\ion{O}{3}]$\lambda$4363.
}\label{figeight}
}
\end{figure}

\begin{figure}
\vbox{
\centerline{
\psfig{figure=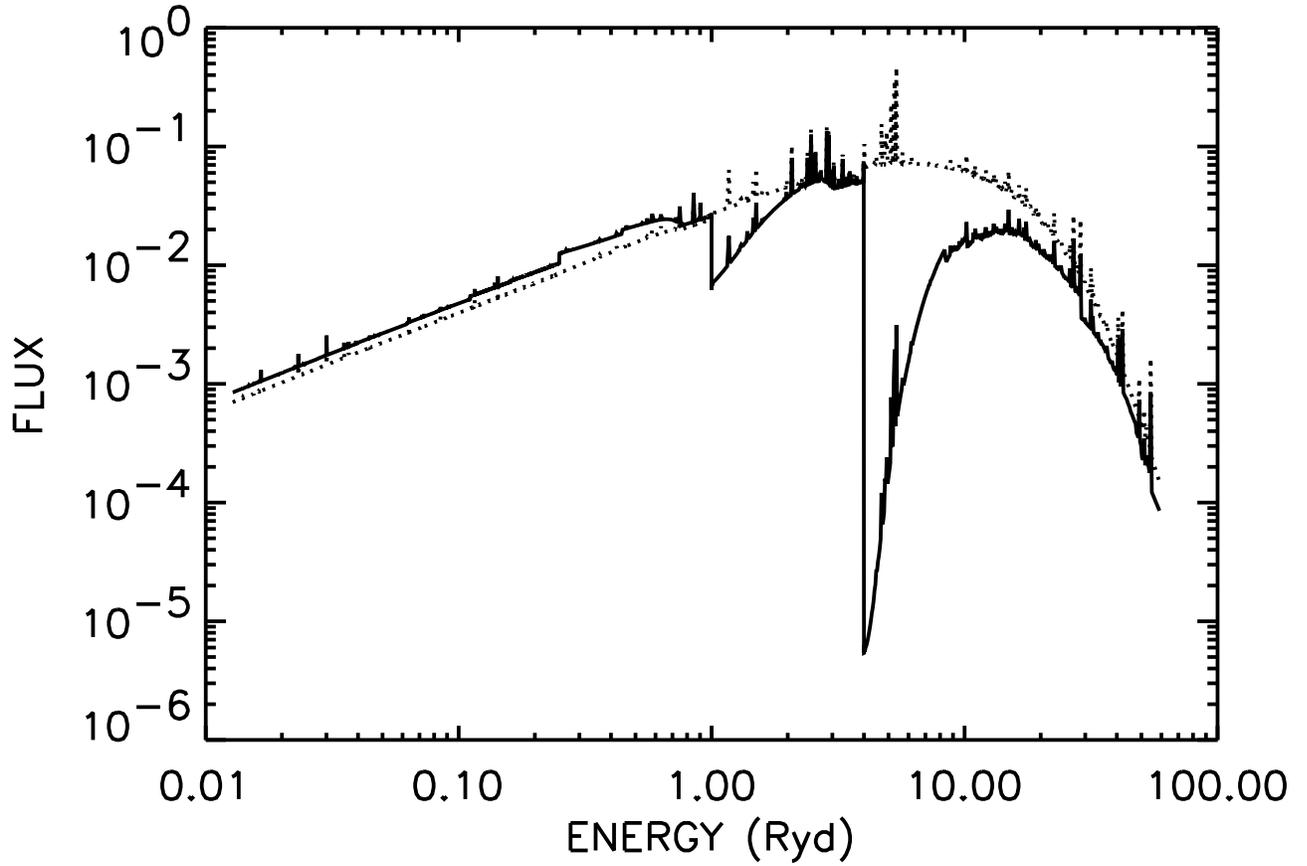,width=7.in}
}
\caption{
Model continua employed in photoionization calculations.
The dotted line is the continuum incident on the screen, while the
solid line is the transmitted continuum incident on the
filament. Free-free, free-bound, and two-photon emission from the
screen make the transmitted continuum higher than the incident one for
energies below $\sim 0.8$ Ryd.
}\label{fignine}
}
\end{figure}

\pagebreak
\begin{table}
\begin{center}
\title{\rm \small Table 1: Observed Emission-Line Fluxes \rm}
\small
\vspace*{0.1in}
\begin{tabular}{ccc}
\tableline
\tableline
Emission Line &Observed Flux$^{1}$ &Dereddened Flux$^{2}$\\
\tableline
$[$\ion{O}{2}$]$$\lambda$3727 & 2.80$\pm$0.10 & 4.80$\pm$0.17   \\
$[$\ion{S}{2}$]$$\lambda$4068 & 0.14$\pm$0.03 &0.19$\pm$0.04\\
H$\delta$ &  0.20$\pm$0.03 &  0.27$\pm$0.04\\
H$\gamma$ &  0.35$\pm$0.04 &  0.42$\pm$0.05\\
$[$\ion{O}{3}$]$$\lambda$4363 & $<$0.035 & $<$0.04    \\ 
\ion{He}{2} $\lambda$4686 & $<$0.02 &$<$0.024\\
H$\beta$ & 1.00$\pm$0.03 &1.00$\pm$0.03 \\
$[$\ion{O}{3}$]$$\lambda$4959 & 0.24$\pm$0.03 & 0.23$\pm$0.03   \\
$[$\ion{O}{3}$]$$\lambda$5007 & 0.66$\pm$0.04 & 0.62$\pm$0.04   \\
$[$\ion{N}{1}$]$$\lambda$5200 &0.33$\pm$0.03 & 0.30$\pm$0.03  \\
\ion{He}{1} $\lambda$5876 & 0.25$\pm$0.02 & 0.19$\pm$0.02 \\
$[$\ion{O}{1}$]$$\lambda$6300 &0.82$\pm$0.04 & 0.59$\pm$0.03\\
$[$\ion{O}{1}$]$$\lambda$6364 &0.33$\pm$0.04 & 0.24$\pm$0.03\\
$[$\ion{N}{2}$]$$\lambda$6548 & 1.22$\pm$0.06 &0.84$\pm$0.04\\
H$\alpha$ & 4.15$\pm$0.14 &2.86$\pm$0.10 \\
$[$\ion{N}{2}$]$$\lambda$6583 & 3.54$\pm$0.12 &2.43$\pm$0.08\\
\ion{He}{1} $\lambda$6678 & 0.07$\pm$0.03 & 0.05$\pm$0.02 \\
$[$\ion{S}{2}$]$$\lambda$6717 &1.38$\pm$0.06 &0.91$\pm$0.04\\
$[$\ion{S}{2}$]$$\lambda$6731 &1.07$\pm$0.05 &0.70$\pm$0.03\\
\tableline
\end{tabular}
\end{center}
$^1$ Normalized to 
H$\beta =  1.93\times10^{-16}$ erg s$^{-1}$ cm$^{-2}$ arcsec$^{-2}$.\\
$^2$ Normalized to 
H$\beta =  7.11\times10^{-16}$ erg s$^{-1}$ cm$^{-2}$ arcsec$^{-2}$, 
after correction for $A_V = 1.2$ mag.\\
\end{table}

\begin{table}
\begin{center}
\title{\rm \small Table 2: Upper Limits on Coronal Lines and 
Theoretical Predictions \rm}
\vspace*{0.1in}
\begin{tabular}{cccc}
\tableline
\tableline
Line/H$\alpha ^1$ & Upper Limit$^{2}$ & Model$^{3}$ &Notes\\
\tableline
$[$\ion{Ni}{12}$]$$\lambda$4232 &1.8$\times 10^{-2}$ &1.2$\times 10^{-4}$ & \\
$[$\ion{Ar}{14}$]$$\lambda$4414 &1.0$\times 10^{-2}$ &1.0$\times 10^{-3}$ &4\\
$[$\ion{Fe}{14}$]$$\lambda$5303 &9.4$\times 10^{-3}$ &1.6$\times 10^{-3}$ &5\\
$[$\ion{Ca}{15}$]$$\lambda$5445 &1.5$\times 10^{-2}$ &0.6$\times 10^{-3}$ & \\
$[$\ion{Fe}{10}$]$$\lambda$6374 &1.4$\times 10^{-2}$ &2.1$\times 10^{-3}$ & \\
$[$\ion{Ni}{15}$]$$\lambda$6702 &7.9$\times 10^{-3}$ &0.1$\times 10^{-3}$ & \\
\tableline
\end{tabular}
\end{center}
$^1$ Surface brightness of H$\alpha$ is 1.9$\times 10^{-15}$~erg~
s$^{-1}$~cm$^{-2}$~arcsec$^{-2}$, after correction for $A_V = 1.2$ mag.\\
$^2$ After correction for $A_V = 1.2$ mag.\\
$^3$ From Sarazin \& Graney (1991), unless otherwise noted.\\
$^4$ From Voit et al. (1994).\\
$^5$ Voit et al. (1994) predict 15.6$\times 10^{-3}$.\\
\end{table}

\begin{table}
\begin{center}
\title{\rm \small Table 3: Line Ratios of the 
Photoionization$+$Screen Model \rm}
\vspace*{0.1in}
\begin{tabular}{ccc}
\tableline
\tableline
Line Ratio &Photoionization Model &Observed Range\\
\tableline
\ion{He}{2} $\lambda$4686/H$\beta$ &0.024 &$<$0.024\\
\ion{He}{1} $\lambda$5876/H$\beta$ &0.18 &0.16-0.20\\
$[$\ion{O}{1}$]$$\lambda$6300/H$\beta$ &0.78 &0.33-0.78\\
$[$\ion{N}{2}$]$$\lambda$6583/H$\beta$ &2.59 &2.10-3.00\\
$[$\ion{S}{2}$]$$\lambda$6717,6731/H$\beta$ &3.00 &1.38-1.83\\
$[$\ion{O}{3}$]$$\lambda$5007/H$\beta$ &0.43 &0.28-0.63\\
\tableline
\end{tabular}
\end{center}
\end{table}


\begin{references}
\reference{} Arabadjis, J. S., \& Bregman, J. N. 1999, \apj, 514, 607
\reference{} Allen, S. W. 1995, \mnras, 276, 947
\reference{} Allen, S. W., Fabian, A. C., Johnstone, R. M., Arnaud, K. E., \& 
Nulsen, P. E. J.  2000, \mnras, submitted  (astro-ph/9910188)
\reference{} Allen, S. W., Fabian, A. C., Nulsen, P. E. J., \& Edge,
A.C. 1992, \mnras, 254, 51
\reference{} Arnaud, K. A., \& Mushotzky, R. F. 1998, \apj, 501, 119
\reference{} Baldwin, J. A., Phillips, M. M., \& Terlevich, R. 
1981, \pasp, 93, 5
\reference{}Begelman, M. C., \& Fabian, A. C. 1990, \mnras, 244,26P
\reference{}Bicknell, G. V., Dopita, M. A., \& O'Dea, C. P. O. 1997, \apj, 
485, 112
\reference{}Binette, L., Dopita, M. A., \& Tuohy, I. R. 1985, \apj, 297, 476
\reference{}Branduardi-Raymont, G., Fabricant, D., Feigelson, E., 
Gorenstein, J., Grindlay, J., Soltan, A., \& Zamorani, G. 1981, \apj, 248, 55
\reference{}Burbidge, E. M., \& Burbidge, G. R. 1965, \apj, 142, 1351
\reference{}Burstein, D., \& Heiles, C. 1984, \apjs, 54, 33
\reference{}Cardelli, J. A., Clayton, G. C., \& Mathis, J. S. 1989, \apj, 
345, 245
\reference{}Cardiel, N., Gorgas, J., \& Aragon-Salamanca, A. 
1995, \mnras, 277, 502
\reference{}Cardiel, N., Gorgas, J., \& Aragon-Salamanca, A. 
1998, \mnras, 298, 977
\reference{}Caulet, A., et al. 1992, \apj, 388, 301
\reference{}Cowie, L. L., Fabian, A. C., \& Nulsen, P. E. J. 1980, \mnras, 
191, 399
\reference{}Crawford, C., \& Fabian, A. C. 1992, \mnras, 259, 265
\reference{}Donahue, M., \& Stocke, J. T. 1994, \apj, 422, 459
\reference{}Donahue, M., \& Voit, G. M. 1991, \apj, 381, 361
\reference{}Donahue, M., \& Voit, G. M. 1993, \apj, 414, L17
\reference{}Dopita, M. A., \& Sutherland, R. S. 1995, \apjs, 102, 161
\reference{}Fabian, A. C., Cowie, L. L., \& Grindlay, J. 1981, \apj, 248, 47
\reference{}Fabian, A. C., Nulsen, P. E. J., \& Canizares, C. R. 1984, \nat, 
310, 733
\reference{}Ferguson, J. W., Korista, K. T., \& Ferland, G. 1997, \apjs, 110, 
287
\reference{}Ferland, G. J. 1996,  Hazy, a Brief Introduction to CLOUDY, 
University of Kentucky Department of Physics and Astronomy Internal 
Report
\reference{}Ferland, G. J., \& Netzer, H. 1983, \apj, 264, 105
\reference{}Filippenko, A. V., \& W. L. W. Sargent 1985, \apjs, 85, 503
\reference{}Filippenko, A. V., \& Terlevich, R. 1992, \apj, 397, L79
\reference{}Friaca, A. C. E., Gonvalves, D. R., Jafelice, L. C., 
Jatenco-Pereira, V., \& Opher, R. 1997, \mnras, 324, 449 
\reference{}Galarza, V. C., Walterbos, R. A. M., \& Braun, R. 1999, \aj, 
118, 2775
\reference{}Ge, J. P., \& Owen, F. N. 1993, \aj, 105, 778
\reference{}Ge, J. P., \& Roberts, J. D. 1993, \baas, 183, 53.10
\reference{}Heckman, T. M. 1980, \aap, 87, 152
\reference{}Heckman, T. M., Armus, L., Weaver, K. A., \& Wang, J. 1999, \apj, 
517, 130
\reference{}Heckman, T. M., Baum, S. A., van Breugel, W. J. M., \& 
McCarthy, P. J. 1989, \apj, 338, 48
\reference{}Held, E. V., Mould, J. R., \& de Zeeuw, P. T. 1990, \aj, 100, 415
\reference{}Holtzman, E. J., et al. 1992, \aj, 103, 691
\reference{}Hoopes, C. G., Walterbos, R. A. M., \& Rand, R. 1999, \apj, 522, 
669
\reference{}Howk, J. C., \& Sembach, K. R. 1999, \apj, 523, L141
\reference{}Hu, M. E., Cowie, L. L., \& Wang, Z. 1985, \apjs, 59, 447
\reference{}Hummer, D. G., \& Storey P. J. 1987, \mnras, 224, 801
\reference{}Jafelice, L. C., \& Friace, A. C. S. 1996, \mnras, 280, 438
\reference{}Johnstone, R. M., \& Fabian, A. C. 1988, \mnras, 233, 581
\reference{}Johnstone, R. M., Fabian, A. C., \& Nulsen, P. E. J. 
1987, \mnras, 224, 75
\reference{}Kent, S. M., \& Sargent, W. L. W. 1979, \apj, 230, 667
\reference{}Kent, S. M., \& Sargent, W. L. W. 1983, \aj, 88, 697
\reference{}Kinney, A. L., Bohlin, R. C., Blades, J. C., \& York, D. G. 
1991, \apjs, 75, 645
\reference{}Kriss, G. 1994, in ASP Conf. Ser. 61, Astronomical Data 
Analysis Software and Systems III, ed. D. R. Crabtree, R. J. 
Hanisch, \& J. Barnes (San Francisco: ASP), 437
\reference{}Lieu, R., Bonamente, M., \& Mittaz, J. P. D. 1999, \apj, 517, L91
\reference{}Lieu, R., Mittaz, J. P. D., Bowyer, S., Lockman, F., \& Hwang, C. 
1996, \apj, 458, L5
\reference{}Lockman, F. J., \& Savage, B. D. 1995, \apjs, 97, 1
\reference{}Martin, C. L. 1997, \apj, 491, 561
\reference{}Martin, C. L., \& Kennicutt, R. C. Jr. 1997, \apj, 483, 698
\reference{}McNamara, B. R., \& O'Connell, R. W. 1989, \aj, 98, 2018
\reference{}McNamara, B. R., O'Connell, R. W., \& Sarazin, C. L. 1996, \aj, 
112, 91
\reference{}McNamara, B. R., et al. 2000, \apj, 534, L135
\reference{}Miller, A. 1997, EUVE Guest Observer Program Handbook, 
4th Ed. (Berkeley, CA: EGO Center)
\reference{}Miller, J. S., \& Stone, R. P. S. 1987, Lick Observatory Technical
Report No. 48, The CCD Cassegrain Spectrograph at the Shane Reflector
(Santa Cruz: Lick Observatory)
\reference{}Minkowski, B. R. 1957, in IAU Symp. 4, Radio Astronomy, ed. 
H. C. van der Hulst (Cambridge: Cambridge Univ. Press), 107
\reference{}Minter, T., \& Spangler, S. R. 1997, \apj, 485, 182
\reference{}Mohan, M., Hibbert, A., \& Kingston, A. E. 1994, \apj, 434, 389
\reference{}Mohr, J. J., Mathiesen, B., \& Evrard, A. E. 1999, \apj, 517
\reference{}Mushotzky, R. F., \& Szymokowiak, A. E. 1988, in Cooling Flows 
in Clusters and Galaxies, ed. A. C. Fabian, (Dordrecht: Kluwer), 53
\reference{}Nelson, C. H., \& Whittle, M. 1995, \apjs, 99, 67
\reference{}Oliva, E. 1997, in ASP Conf. Ser. 113, Emission Lines in 
Active Galaxies: New Methods and Techniques, ed. B. M. Peterson, F. Cheng, \& 
A. S. Wilson (San Francisco: ASP), 228
\reference{}Osterbrock, D. E. 1989, Astrophysics of Gaseous Nebulae and Active 
Galactic Nuclei (Mill Valley, CA: University Science Books)
\reference{}Pedlar, A., Booler, V., \& Davies, R. V. 1983, \mnras, 203, 667
\reference{}Pedlar, A., Ghantaure, H. S., Davies, R. V., Harrison, B. A., 
Perley, R., Crane, P. C., \& Unger, S. W. 1990, \mnras, 246, 477
\reference{}Pelan, J., \& Berrington, K. A. 1995, \aap, 110, 209
\reference{}Pringle, J. E. 1989, \mnras, 239, 479
\reference{}Rand, R. 1998, \apj, 501, 137
\reference{}Raymond, J. C. 1979, \apjs, 39, 1
\reference{}Reynolds, R. J. 1985, \apj, 294, 256
\reference{}Romanishin, W. 1987, \apj, 323, L113
\reference{}Rubin, V. C., Ford, W. K., Peterson, C. J., \& Lynds, C. R. 
1978, \apjs, 37, 235
\reference{}Sarazin, C. L. 1999, \apj, 520, 529
\reference{}Sarazin, C. L., \& Lieu, R. 1998, \apj, 494, L177
\reference{}Sarazin, C. L., \& Graney, C. M. 1991, \apj, 375, 532
\reference{}Sembach, K. R., Howk, J. C., Ryans, R. S. I., \& Keenan, F. P. 
2000, \apj, 528, 310
\reference{}Seyfert, C. A., 1943, \apj, 97, 28
\reference{}Shaw, R. A., \& Dufour, R. J. 1995, \pasp, 107, 896
\reference{}Shields, J. C., 1992, \apj, 399, L27
\reference{}Shields, J. C., \& Filippenko, A. V. 1990a, \apj, 353, L7
\reference{}Shields, J. C., \& Filippenko, A. V. 1990b, \aj, 100, 1034
\reference{}Shopbell, P. L., \& Bland-Hawthorn, J. 1998, \apj, 493, 129
\reference{}Shull, J. M., \& McKee, C. 1979, \apj, 227, 131
\reference{}Sparks, W. 1992, \apj, 399, 66
\reference{}Storey, P. J., Mason, H. E., \& Saraph, H. E., 1996, \aap, 309, 677
\reference{}Soker, N., \& Sarazin, C. L. 1990, \apj, 348, 73
\reference{}Taylor, G. B., Barton, E. J., \& Ge, J. P. 1994, \aj, 107, 1942
\reference{}Unger, S. W., Taylor, K., Pedlar, A., Ghantaure, H. S., 
Penston, M. V., \& Robinson, A. 1990, \mnras, 242, 33
\reference{}van Breugel, W., Heckman, T. M., \& Miley, G. 1984, \apj, 276, 79
\reference{}Veilleux, S., \& Osterbrock, D. E. 1987, \apjs, 63, 295
\reference{}Voit, G. M., \& Donahue, M. 1997, \apj, 486, 242
\reference{}Voit, G. M., \& Donahue, M. 1990, \apj, 360, L15
\reference{}Voit, G. M., \& Donahue, M., \& Slavin, J. D. 1994, \apjs, 95, 87
\reference{}Wirth, A., Kenyon, S. J., \& Hunter, D. A. 1983, \apj, 269, 102
\reference{}Yan, L., \& Cohen, J. G. 1995, \apj, 454, 44
\end{references}
\end{document}